\newif\ifcheckpagelimits
\checkpagelimitstrue
\checkpagelimitsfalse

\ifcheckpagelimits
 \documentclass[nofootinbib,pra,aps,twocolumn,showpacs,showkeys,%
 amsmath,amssymb,superscriptaddress,final,reprint,floatfix,longbibliography]{revtex4-1}
 \newcommand{\todo}[1]{}
\else
 \documentclass[pra,aps,twocolumn,showpacs,showkeys,%
 amsmath,amssymb,superscriptaddress,final,reprint,floatfix,longbibliography]{revtex4-1}
 \newcommand{\todo}[1]{{\pdfmargincomment[icon=Note,color=pink]{#1}}}
\fi

\usepackage{amssymb}

\newcommand{\vewedgebullet}{\mathrel{ \veebar \!\!\!\!\!\!\!\phantom{.}_\bullet}}

\newcommand{\barwedgebullet}{ \mathrel{\scalebox{0.72}{\reflectbox{\rotatebox[origin=c]{180}{$\vewedgebullet$}}}}}

\usepackage[table]{xcolor}
\usepackage{helvet} 
\usepackage{lineno}

  \usepackage{mathptmx}

\usepackage{subfigure}
\usepackage{hhline} 
\usepackage{psfrag,graphicx}
\usepackage{dcolumn}
\usepackage{amsmath,amssymb}
\usepackage{bm}
\usepackage{color}
\usepackage{overpic}
\usepackage{latexsym}
\usepackage{epstopdf}
\usepackage{color}
\usepackage[english]{babel}
\usepackage{latexsym}
\usepackage{stmaryrd}

\usepackage{braket}
\usepackage{tikz}
\usetikzlibrary{shapes.geometric}
\newcommand{\tboxed}[2][inner sep=1pt]{\ifmmode
\tikz[baseline=(X.base),outer
sep=0pt]{\node[draw,#1](X){\ensuremath{#2}};}
\else
\tikz[baseline=(X.base),outer
sep=0pt]{\node[draw,#1](X){#2};}
\fi
}
\newcommand{\circled}[2][inner sep=1pt]{\ifmmode
\tikz[baseline=(X.base),outer sep=0pt]{\node[circle,draw,#1](X){\ensuremath{#2}};}
\else
\tikz[baseline=(X.base),outer sep=0pt]{\node[circle,draw,#1](X){#2};}
\fi
}
\newcommand{\rhombed}[2][inner sep=1pt]{\ifmmode
\tikz[baseline=(X.base),outer sep=0pt]{\node[diamond,draw,#1](X){\ensuremath{#2}};}
\else
\tikz[baseline=(X.base),outer sep=0pt]{\node[diamond,draw,#1](X){#2};}
\fi
}
\newcommand{\trapeziumR}[2][inner sep=0pt]{\ifmmode
\tikz[baseline=(X.base),outer sep=0pt]{\node[trapezium,draw,
trapezium left angle=120,trapezium right angle=60,#1](X){\ensuremath{#2}};}
\else
\tikz[baseline=(X.base),outer sep=0pt]{\node[trapezium,draw,
trapezium left angle=120,trapezium right angle=60,#1](X){#2};}
\fi
}
\newcommand{\trapeziumL}[2][inner sep=0pt]{\ifmmode
\tikz[baseline=(X.base),outer sep=0pt]{\node[trapezium,draw,
trapezium left angle=60,trapezium right angle=120,#1](X){\ensuremath{#2}};}
\else
\tikz[baseline=(X.base),outer sep=0pt]{\node[trapezium,draw,
trapezium left angle=60,trapezium right angle=120,#1](X){#2};}
\fi
}

\definecolor{mygrey}{gray}{0.94}
\definecolor{myblue}{rgb}{0.2,0.2,0.8}
\definecolor{myzard}{cmyk}{0,0,0.05,0}
\definecolor{mywhite}{rgb}{1,1,1}
\definecolor{myred}{rgb}{1,0.,0.3}

\usepackage[colorlinks=true,citecolor=myblue,linkcolor=myred]{hyperref}
\DeclareMathAlphabet{\mathpzc}{OT1}{pzc}{m}{it}

 \def\ee{\mathord{\rm e}}
 
 \def\ii{\mathord{\rm i}}
 \def\mod{\mathord{\rm mod}}

\renewcommand{\ii}{{\rm i}}
\renewcommand{\ee}{{\rm e}}
\def\beq{\begin{equation}}
\def\eeq{\end{equation}}

\def\barray{\begin{eqnarray}}
\def\earray{\end{eqnarray}}

\definecolor{LightGrey}{gray}{0.85}
\definecolor{LightGreen}{rgb}{0.60,1,0.2}
\definecolor{UberLightGreen}{rgb}{0.80,1,0.6}
\definecolor{LightOrange}{rgb}{1,1,0.39}

\newlength\figureheight
\newlength\figurewidth

\def\ket#1{\left|#1\right>}
\def\bra#1{\left<#1\right|}

 \def\ee{\mathord{\rm e}}
 
 \def\ii{\mathord{\rm i}}
 \def\mod{\mathord{\rm mod}}

\renewcommand{\ii}{{\rm i}}
\renewcommand{\ee}{{\rm e}}

\definecolor{mygrey}{gray}{0.94}
\definecolor{myred}{rgb}{1,0.,0.3}

\newcommand{\plaquetteA}{\mathrel{ \trapeziumL{\phantom{:}}\hspace{-2.1ex}_\bullet}}

\newcommand{\plaquetteB}{ \mathrel{\scalebox{0.97}{\reflectbox{\rotatebox[origin=c]{180}{$\plaquetteA$}}}}}

\begin{document}


\title{Robust topological order in fermionic $\mathbb{Z}_2$  gauge theories: \\ from Aharonov-Bohm instability to soliton-induced deconfinement}


\author{Daniel Gonz\'{a}lez-Cuadra}\email{daniel.gonzalez@icfo.eu}
\affiliation{ICFO - Institut de Ci\`encies Fot\`oniques, The Barcelona Institute of Science and Technology, Av. Carl Friedrich Gauss 3, 08860 Castelldefels (Barcelona), Spain}

\author{Luca Tagliacozzo}
\affiliation{Departament de F\'{i}sica Qu\`antica i Astrof\'{i}sica and Institut de Ci\`encies del Cosmos (ICCUB), Universitat de Barcelona, Mart\' {i} i Franqu\`es 1, 08028 Barcelona, Spain}

\author{Maciej Lewenstein}
\affiliation{ICFO - Institut de Ci\`encies Fot\`oniques, The Barcelona Institute of Science and Technology, Av. Carl Friedrich Gauss 3, 08860 Castelldefels (Barcelona), Spain} 
\affiliation{ICREA, Lluis Companys 23, 08010 Barcelona, Spain}

\author{Alejandro Bermudez}
\affiliation{Departamento de F\'{i}sica Te\'{o}rica, Universidad Complutense, 28040 Madrid, Spain}

\begin{abstract}
Topologically-ordered phases of matter, although stable against local perturbations, are usually restricted to relatively small regions in phase diagrams. Their preparation requires thus a precise fine tunning of the system's parameters, a very challenging task in most experimental setups. In this work, we investigate a model of spinless fermions interacting with dynamical $\mathbb{Z}_2$ gauge fields on a cross-linked ladder, and show evidence of topological order throughout the full parameter space. In particular, we show how a magnetic flux is spontaneously generated through the ladder due to an Aharonov-Bohm instability, giving rise to topological order even in the absence of a plaquette term. Moreover, the latter coexists here with a symmetry-protected topological phase in the matter sector, that displays fractionalised gauge-matter edge states, and intertwines with it by a flux-threading phenomenon. Finally, we unveil the robustness of these features through a gauge frustration mechanism, akin to geometric frustration in spin liquids, allowing topological order to survive to arbitrarily large quantum fluctuations. In particular, we show how, at finite chemical potential, topological solitons are created in the gauge field configuration, which bound to fermions forming $\mathbb{Z}_2$ deconfined quasi-particles. The simplicity of the model makes it an ideal candidate where 2D gauge theory phenomena, as well as exotic topological effects, can be investigated using cold-atom quantum simulators.

\end{abstract}

\maketitle

\setcounter{tocdepth}{2}
\begingroup
\hypersetup{linkcolor=black}
\tableofcontents
\endgroup

\section{\bf Introduction}
Understanding  \emph{quantum many-body  systems} is generally a hard problem, as their complexity increases exponentially with the number of constituents. From this large complexity, exotic collective phenomena may arise,    as occurs for the  so-called spin liquids. These phases of matter    evade spontaneous symmetry breaking, and thus long-range order~\cite{landau_symm_breaking}, down to the lowest possible temperatures~\cite{anderson_1987,misguich_2008,balents_2010}. In spite of this, spin liquids can be   characterized by a different notion of order: \emph{topological order}~\cite{RevModPhys.89.041004}. Systems with topological order have degenerate ground-states, the number of which depends on the underlying topology. Each ground-state is a strongly-correlated state, as witnessed by the multipartite long-range entanglement among the constituents~\cite{kitaev_2006,levin_2006}. 
Besides, the ground-state manifold is separated from the rest of the spectrum by a finite energy gap and, more importantly, only non-local perturbations can act non-trivially within it. It is thus  a natural subspace to encode  quantum information, and a promising route for   fault-tolerant quantum computers~\cite{kitaev_2003,Nayak_2008}. 

Unfortunately, topological order is very elusive and tends to be fragile, as witnessed by the few materials where it has been observed, requiring in many cases extremely low temperatures and very high purity in the samples \cite{han_2012,wen_2019}. Here we identify a promising route to prepare robust topologically-ordered states in cold-atom systems using gauge invariance.

Gauge theories,  used to describe strong, weak and electromagnetic interactions~\cite{kogut_1979}, have local symmetries that cannot be broken spontaneously~\cite{PhysRevD.12.3978}, evading thus the standard form of ordering. For this reason~\cite{fradkin_2013}, emergent gauge theories   also play an important role in  long-wavelength descriptions  of  non-standard phases of matter, such as high-$T_{\rm c}$ superconductors~\cite{PhysRevB.37.580} and frustrated magnets~\cite{PhysRevLett.66.1773}. Formally, gauge theories can be described through  Hamiltonians  that commute with an extensive number of local symmetry operators forming a group, the gauge group~\cite{PhysRevD.11.395}.

Pure gauge theories describe the physics of gauge bosons, the generalization of photons to arbitrary gauge groups~\cite{PhysRev.96.191}, and host different phases that can be  characterized by the potential that the bosons  mediate between  test charges~\cite{kogut_1979,greensite_2011}. In a \emph{deconfined phase},    particles generated in pairs of opposite charge can be  separated arbitrarily far away with a finite energy cost. Conversely, there can also exist confined phases where this potential energy increases linearly with the distance. The simplest gauge theory on the lattice~\cite{PhysRevD.10.2445}, the so-called $\mathbb{Z}_2$ or Ising gauge theory (IGT),  already gives rise to  a confined-deconfined phase transition without spontaneous symmetry breaking~\cite{Wegner_z2}. We note that the very nature of this deconfined  phase  is the key underlying Kitaev's toric code~\cite{kitaev_2003}, a spin-liquid phase allowing for topological quantum error correction and fault-tolerant quantum computing~\cite{RevModPhys.87.307}.  It is thus important,  both from fundamental  and applied perspectives, to study the fate of the IGT deconfined phase and, more generally, its full phase diagram as perturbations are introduced~ \cite{kitaev_2003,trebst_2007,hamma_2008,vidal_2009,tagliacozzo_2011}. 
Understanding such phase diagrams when the gauge fields interact with matter fields, either bosonic or fermionic, is generally a very hard problem with longstanding open questions~\cite{kogut_1979}. In the simplest case, the topologically-ordered deconfined phase of  the IGT
coupled to dynamical $\mathbb{Z}_2$ matter can be understood  through the toric code perturbed by both parallel and transverse fields~\cite{fradkin_1979,kitaev_2003,vidal_2009a,tupitsyn_2010}. While the corresponding phase diagram  is known since the late 70s~\cite{fradkin_1979}, exchanging $\mathbb{Z}_2$ for  fermionic matter leads to a much richer scenario, which is only beginning to be explored~\cite{Assaad_2016,Gazit_2017,Prosko_2017,Gazit_2018,1912.11106}.

These connections have fuelled a multi-disciplinary effort towards, not only improving our understanding of these lattice gauge theories (LGTs), but also  realizing them experimentally, either in natural or in synthetic quantum materials, such as cold atoms in optical lattices~\cite{lewenstein_2012}. These are systems where atoms are very dilute and, thus, primarily interact by $s$-wave scattering. Trapping the atoms by an optical lattice allows to reach the strongly-interacting regime, but the interactions are still limited to be on-site~\cite{bloch_2008}. This fact constitutes a major hurdle when trying to realize  lattice gauge theories with ultra-cold atoms~\cite{1911.00003,Wiese_2013, Zohar_2015, Dalmonte_2016}, as they require interactions between all the atoms connected through elementary loops of the lattice (i.e. plaquettes)~\cite{PhysRevLett.95.040402, PhysRevA.73.022328,Zohar_2011, Zohar_2012, PhysRevLett.110.125304, PhysRevLett.110.125303,Tagliacozzo2013, Tagliacozzo_2013, Zohar_2017}. Aside from this point, the implementation of the tunnelling of matter dressed by the gauge fields is also far from trivial.  Floquet engineering in strongly-interacting gases~\cite{bermudez_2015,dutta_2017,Barbieroeaav7444}, and spin-changing collisions in atomic mixtures~\cite{PhysRevA.88.023617,Kasper_2017,abelian_higgs,PhysRevB.100.115152}, have identified neat directions towards this goal, which are particularly promising in light of recent  experiments~\cite{Schweizer_2019, Gorg_2019, Mil_2019, Yang_2020}. Since the realization of plaquette terms is currently the major experimental bottleneck to simulate gauge theories beyond 1D,  a timely question would be: is it possible to find characteristic features, such as deconfinement and topological order, in lattice gauge theories without plaquette terms?

\begin{figure}[t]
  \centering
  \includegraphics[width=1.0\linewidth]{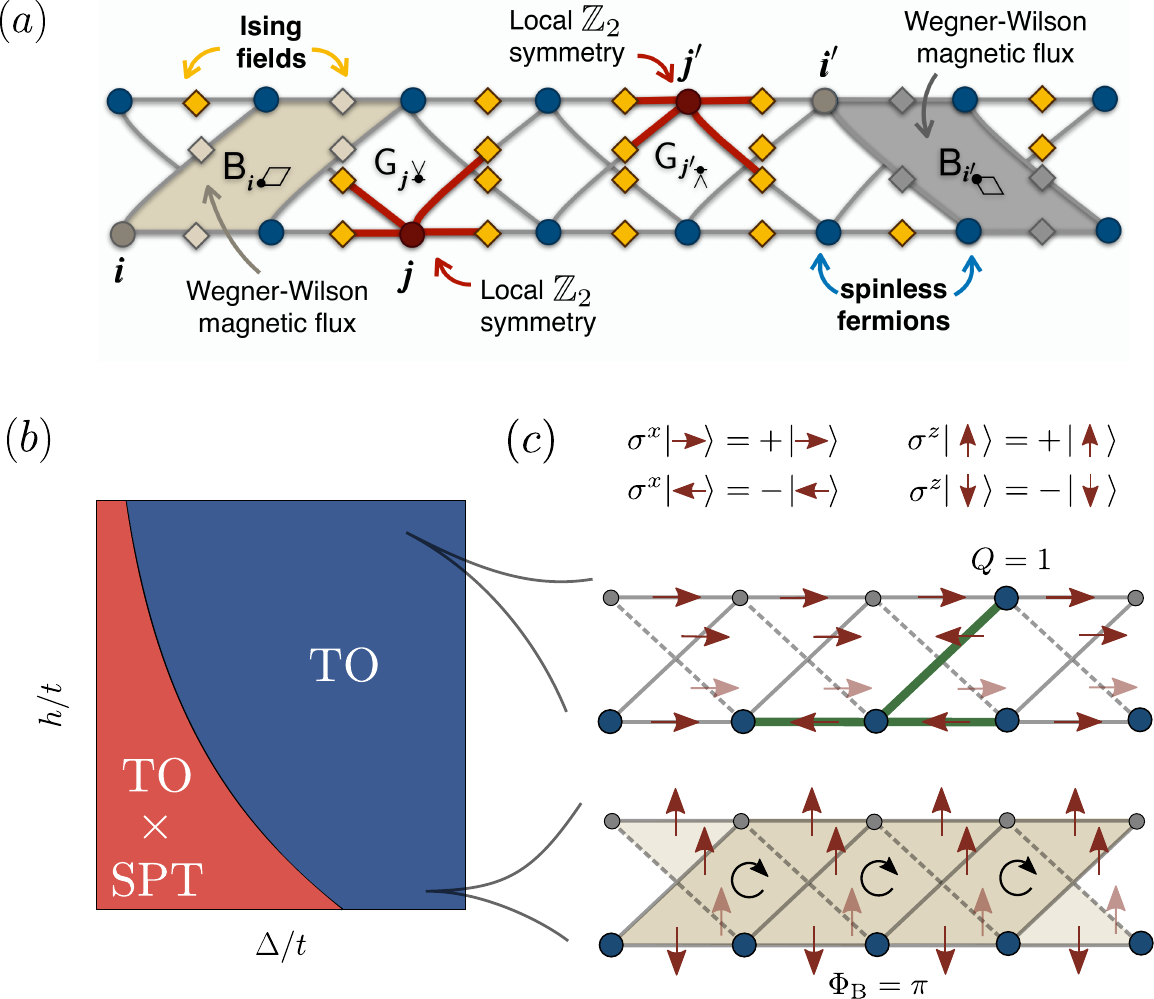}
\caption{\label{fig:scheme} \textbf{The Creutz-Ising ladder:} \textbf{(a)}  Spinless fermions reside on the sites of a two-leg ladder (filled circles), and can tunnel along and across the legs forming a cross-linked pattern. These tunnelings are minimally dressed by Ising spin-$1/2$ fields, which sit on the corresponding links (filled rhombi). We represent the Wegner-Wilson fluxes across the two minimal plaquettes (titled trapezes in grey), and the generators of the local $\mathbb{Z}_2$ symmetries (red graphs). \textbf{(b)} Sketch of the phase diagram at half filling in terms of the electric field $h$ and the imbalance $\Delta$. We found two phases, both with topological order (TO). In one case, the latter coexists with a symmetry-protected topological phase (SPT) in the matter sector. \textbf{(c)} For $\Delta \gg t$, the fermions populate the lower leg of the ladder. At $h \ll t$, the Ising fields rearranged to spontaneously generate a $\pi$ flux per plaquette through an Aharonov-Bohm instability, giving rise to TO. For $h \gg t$, the latter survives due to a gauge frustration mechanism. Fermions delocalise by forming bound quasiparticles with topological defects created in an otherwise dimerized electric field background.}
\end{figure}

In this work, we  show that this is indeed possible. By studying a cross-linked  lattice connectivity (see Fig.~\ref{fig:scheme}), we  identify a  new avenue for the interplay of local symmetries and topology in LGTs, as an Aharonov-Bohm instability can induce a magnetic flux in the absence of plaquette terms, giving rise to topological order, that, in this case, coexists with a symmetry-protected topological (SPT) phase~\cite{spt_phases_review}. The crucial role that gauge symmetry plays in the topological properties of the system extends to large quantum fluctuations through a frustration mechanism, allowing deconfinement to survive to the whole phase diagram.

The paper is organized as follows. In Sec.~\ref{sec:model}, we introduce the {\it Creutz-Ising ladder}, a quasi-1D $\mathbb{Z}_2$ LGT where the Ising fields are coupled to spinless fermions hopping in a cross-linked ladder, and summarise our main findings. In Sec.~\ref{sec:instability}, we describe the Aharonov-Bohm instability and the emergence of a magnetic flux, which gives rise to an SPT phase. We study this phenomenon in the presence of quantum fluctuations of the gauge fields, and provide a full discussion of the phase diagram. In Sec.~\ref{sec:topology}, we demonstrate that  the cross-linked ladder can be understood as the thin-cylinder limit of a 2D LGT, providing a practical scenario where the ground-state degeneracy is related to the topology of the underlying manifold. In Sec.~\ref{sec:deconfinement}, we explore the mechanism of fermionic deconfinement mediated by topological solitons, which can be neatly understood in the limit of large quantum fluctuations through a gauge frustration effect. Finally, we present our conclusions and outlook in Sec.~\ref{sec:conclusions}.

\section{\bf The  Creutz-Ising ladder}
\label{sec:model}

\subsection{The model} 
The  Creutz ladder, which describes  spinless fermions  on a cross-linked ladder~\cite{PhysRevLett.83.2636}, is a  lattice model hosting an  SPT phase. The  tunnelling of fermions   is  dressed by a static  magnetic field that pierces the ladder, which is described by a gauge-invariant flux that pierces the elementary plaquettes. For a static $\pi$ flux, the ground-state of this model may correspond to either the $\mathsf{BDI}$, or the $\mathsf{AIII}$ class of topological insulators~\cite{RevModPhys.88.035005}, a free-fermion  insulating SPT phase. Interestingly,  the physics of cross-linked ladders has already been explored in experiments of ultracold atoms by exploiting Floquet engineering in  two-orbital optical lattices~\cite{PhysRevLett.121.150403,Kang_2020}. To go beyond this  free-fermion scenario, a natural possibility is to include Hubbard-type interactions~\cite{PhysRevX.7.031057}, which leads to correlated SPT phases with interesting connections to relativistic quantum field theories of  self-interacting fermions~\cite{BERMUDEZ2018149,PhysRevB.99.125106}.

We hereby follow a different, and yet unexplored, route: we upgrade the background magnetic fields  to a $\mathbb{Z}_2$ LGT by introducing Ising fields on the links (see Fig. \ref{fig:scheme}\textbf{(a)}). This IGT is  described by the following Hamiltonian
\beq
\label{eq:H_CI}
\mathsf{H}_{\rm CI}(t,\Delta,h)=\sum_{\boldsymbol{i}}\sum_{( \boldsymbol{i},\boldsymbol{j})}\left(-tc^\dagger_{\boldsymbol{i}}\sigma^z_{(\boldsymbol{i},\boldsymbol{j})}c^{\phantom{\dagger}}_{\boldsymbol{j}}-h\sigma^x_{(\boldsymbol{i},\boldsymbol{j})}\right)+\frac{\Delta}{2}\sum_{ \boldsymbol{i}}  s^{\phantom{\dagger}}_{\boldsymbol{i}}c^\dagger_{\boldsymbol{i}}c^{\phantom{\dagger}}_{\boldsymbol{i}},
\eeq 
where $c^\dagger_{\boldsymbol{i}} (c^{\phantom{\dagger}}_{\boldsymbol{i}})$  creates (annihilates)  a  fermion  at   site $\boldsymbol{i}=(i_1,i_2)$. Here,  $i_2\in\mathbb{Z}_2=\{0,1\}$ labels the lower and upper legs of the ladder, and $i_1\in\mathbb{Z}_{N_{\rm s}}=\{0,\cdots,N_{\rm s}-1\}$ labels the sites of each of these legs.    At the   horizontal or diagonal  links  $(\boldsymbol{i},\boldsymbol{j})$ adjacent to $\boldsymbol{i}$, we introduce the Pauli matrices  $\sigma^z_{(\boldsymbol{i},\boldsymbol{j})}, \sigma^x_{(\boldsymbol{i},\boldsymbol{j})}$ as the corresponding Ising link operators.  The first term of Eq.~\eqref{eq:H_CI} describes  the tunnelling of fermions dressed by the Ising gauge fields, which has  tunnelling strength $t$. The second term introduces an electric transverse field of strength $h$. Finally, the third term describes an energy imbalance of magnitude $\Delta$ for the fermions sitting on the upper  $s_{\boldsymbol{i}}=+1$,or lower  leg $s_{\boldsymbol{i}}=- 1$.

The above Hamiltonian~\eqref{eq:H_CI} displays a local $\mathbb{Z}_2$ symmetry $[\mathsf{H}_{\rm CI},\mathsf{G}_{\boldsymbol{i}}]=0,\forall \boldsymbol{i}\in\mathbb{Z}_{N_{\rm s}}\times\mathbb{Z}_{2}$, with the  generators 
\beq
\label{eq:generators}
\mathsf{G}_{_{\boldsymbol{i}}\vewedgebullet}=(-1)^{c^\dagger_{\boldsymbol{i}}c^{\phantom{\dagger}}_{\boldsymbol{i}}}\!\!\!\!\prod_{(\boldsymbol{i},\boldsymbol{j})\in_{\boldsymbol{i}}\vewedgebullet}\!\sigma^x_{(\boldsymbol{i},\boldsymbol{j})},\hspace{3ex} \mathsf{G}_{^{\boldsymbol{i}}\barwedgebullet}=(-1)^{c^\dagger_{\boldsymbol{i}}c^{\phantom{\dagger}}_{\boldsymbol{i}}}\!\!\!\!\prod_{(\boldsymbol{i},\boldsymbol{j})\in ^{\boldsymbol{i}}\barwedgebullet}\!\sigma^x_{(\boldsymbol{i},\boldsymbol{j})},
\eeq 
 displayed in Fig.~\ref{fig:scheme}\textbf{(a)}. In addition, we also depict in this figure the smallest 
Wegner-Wilson loops, corresponding  to gauge-invariant magnetic fields across two types of trapezia
\beq
\label{eq:four-body_flux}
\textsf{B}_{_{\boldsymbol{i}}\hspace{0.2ex}\plaquetteA}\hspace{1.5ex}=\prod_{(\boldsymbol{i},\boldsymbol{j})\in _{\boldsymbol{i}}\hspace{0.2ex}\plaquetteA }\sigma^z_{(\boldsymbol{i},\boldsymbol{j})},\hspace{3ex} \textsf{B}_{^{\boldsymbol{i}}\hspace{0.2ex}\plaquetteB}\hspace{1.5ex}=\prod_{(\boldsymbol{i},\boldsymbol{j})\in\hspace{0.1ex}
^{\boldsymbol{i}}\hspace{0.2ex}\plaquetteB}\sigma^z_{(\boldsymbol{i},\boldsymbol{j})},
\eeq
with corresponding magnetic fluxes
\beq
\Phi_{\textsf{B}}^{_{\boldsymbol{i}}\hspace{0.2ex}\plaquetteA}\hspace{1ex}=\arccos(\langle\textsf{B}_{_{\boldsymbol{i}}\hspace{0.2ex}\plaquetteA}\hspace{1.5ex}\rangle),\hspace{3ex}\Phi_{\textsf{B}}^{_{\boldsymbol{i}}\hspace{0.2ex}\plaquetteB}\hspace{1ex}=\arccos(\langle\textsf{B}_{_{\boldsymbol{i}}\hspace{0.2ex}\plaquetteB}\hspace{1.5ex}\rangle).
\eeq
The magnetic flux that threads a plaquette corresponds to the phase accumulated by a particle that encircles that plaquette. Using this picture, we can write down the spin operators present in the gauge-invariant tunneling terms of \eqref{eq:H_CI} as dynamical $\mathbb{Z}_2$ Peierls phases, $\sigma^z_{(\boldsymbol{i},\boldsymbol{j})} = e^{i\varphi_{(\boldsymbol{i},\boldsymbol{j})}}$, where $\varphi_{(\boldsymbol{i},\boldsymbol{j})}$ has eigenvalues $0$ and $\pi$.

 We note that, in the standard formulation of IGTs~\cite{RevModPhys.51.659}, one also introduces an additional  magnetic-field term  
 \beq
 \label{eq:full_ham}
 \tilde{\mathsf{H}}_{\rm CI}(t,\Delta,h,J)=\mathsf{H}_{\rm CI}(t,\Delta,h)-J\sum_{\boldsymbol{i}}\left(\textsf{B}_{_{\boldsymbol{i}}\hspace{0.2ex}\plaquetteA}\hspace{1.5ex}+\textsf{B}_{^{\boldsymbol{i}}\hspace{0.2ex}\plaquetteB}\hspace{1.ex}\right),
 \eeq
  such that the magnetic plaquette coupling  $J$ competes with the electric transverse field $h$. In the (2+1) pure IGT, this competition leads to a quantum phase transition between  deconfined  $h/J<h/J|_{\rm c}$ and  confined  $h/J>h/J|_{\rm c}$ phases~\cite{Wegner_z2}. These phases are not characterised by a local order parameter, but instead display  Wegner-Wilson loops that scale either with the perimeter ($h/J<h/J|_{\rm c}$) or with the encirled area  ($h/J<h/J|_{\rm c}$) of a closed loop, i.e. perimeter or area law.
 
 The $\mathbb{Z}_2$ symmetry generators~\eqref{eq:generators} can be used to define different charge sectors of the Hilbert space, as the eigenstates of the Hamiltonian $\ket{\psi}$ must also fulfill 
 \beq
 \label{eq:gauss_law}
 \mathsf{G}_{\boldsymbol{i}}\ket{\psi}=(-1)^{q_{\boldsymbol{i}}}\ket{\psi},
 \eeq
  where $q_{\boldsymbol{i}}\in\{0,1\}$ are the so-called  static $\mathbb{Z}_2$ charges. Typically, one considers the vacuum/even sector $\{q_{\boldsymbol{i}}\}=\{0,0,\cdots,0\}$, introducing  a few static charges on top of it. For instance, $\{q_{\boldsymbol{i}}\}=\{\delta_{\boldsymbol{i},(i_0,0)},\delta_{\boldsymbol{i},({i}_0+L,0)} \}$ describes a pair of  static $\mathbb{Z}_2$  charges separated by a distance $L$. In the (2+1) pure IGT~\cite{kogut_1979}, these test charges are subjected to a potential $V(L)=E_{\rm gs}(L)-E_{\rm gs}(0)$ that  either remains constant in the deconfined phase $V(L)\propto V_0$, or increases  with the distance  in the confined phase $V(L)\propto L$. We note that (2+1) is the lower critical dimension, since  the (1+1) IGT can only display an area law~\cite{kogut_1979}, hosting solely a  confined phase. In the presence of fermionic matter, rather than through the aforementioned area law, the (1+1) confined phase can be characterised through the appearance of chargeless bound dimers~\cite{grudst_1d_z2}.
  
  In this work, we argue that fermionic $\mathbb{Z}_2$ gauge theories in quasi-1D geometries, such as the ladder structure of Fig.~\ref{fig:scheme},  lead to a much richer playground in comparison to the strict 1D limit (Fig. \ref{fig:scheme}\textbf{(b)}). Let us summarise our main findings. 
 
\subsection{Summary of our results}
\label{sec:summary}
 
 In the pure gauge sector, which is obtained from Eq.~\eqref{eq:full_ham} by setting $t=\Delta=0$, we show that $\tilde{\mathsf{H}}_{\rm CI}(0,0,h,J)$ still hosts  a  quantum phase transition  at a critical $h/J|_{\rm c}$, separating confined and deconfined phases. We characterise this  phase transition  quantitatively using matrix-product-state (MPS) numerical simulations~\cite{SCHOLLWOCK201196}, which allow us to extract the critical behaviour of the Ising magnetic fluxes, and their susceptibilities. We note that the ladder geometry plays a key role to go beyond the (1+1) lower critical dimension~\cite{kogut_1979}. By switching on the coupling to the dynamical fermions, we show that the aforementioned  Aharonov-Bohm instability  takes place, and results in an emerging $\pi$-flux deconfined phase even in the absence of the magnetic plaquette term (Fig. \ref{fig:scheme}\textbf{(c)}), namely setting
$J=0$ in  the Creutz-Ising Hamiltonian $\tilde{\mathsf{H}}_{\rm CI}(t,\Delta,h,0)$.  We explicitly demonstrate   the presence of topological order  by calculating the topological entanglement entropy associated to the ground-state wavefunction~\cite{PhysRevLett.96.110404,PhysRevLett.96.110405}. 
As opposed to the pure gauge theory, we show how the accompanying deconfinement survives in the limit of arbitrary quantum fluctuations set by  large transverse fields $h$ (Fig. \ref{fig:scheme}\textbf{(c)}). Here, single $\mathbb{Z}_2$ charges can be localised within topological solitons that interpolate between two different symmetry-breaking orders. We believe that this is a generic feature of IGTs in the particular charge sector considered in this work. To the best of our knowledge, this study provides the first quantitative analysis of such deconfinement mechanism.

 Moreover, as a result of the cross-linked geometry, we also show that the matter sector may lie in an SPT phase characterised by a non-zero topological invariant. From the perspective of the fermions,  the corresponding topological edge states can be understood as domain-wall fermions~\cite{KAPLAN1992342,JANSEN1992374,GOLTERMAN1993219} with the novelty that, instead of requiring fine tuning to incorporate chiral symmetry on the lattice,  they are spontaneously generated by the Ising-matter coupling. The fact that a plaquette term $J\neq0$ is not required to host this exotic behaviour is particularly interesting in light of current developments in cold-atom quantum simulations.  Interestingly, the interplay between geometry and gauge-invariant interactions allows us to  obtain a topological phase for gauge fields without introducing four-body plaquette terms, simplifying enormously the experimental implementation. This point is  important since the main building blocks of the model have already been realized in cold-atom experiments~\cite{Schweizer_2019}. Therefore, future quantum simulations of this fermionic IGT will be capable of testing the non-trivial equilibrium properties described in this work.

\begin{figure}[t]
  \centering
  \includegraphics[width=1.0\linewidth]{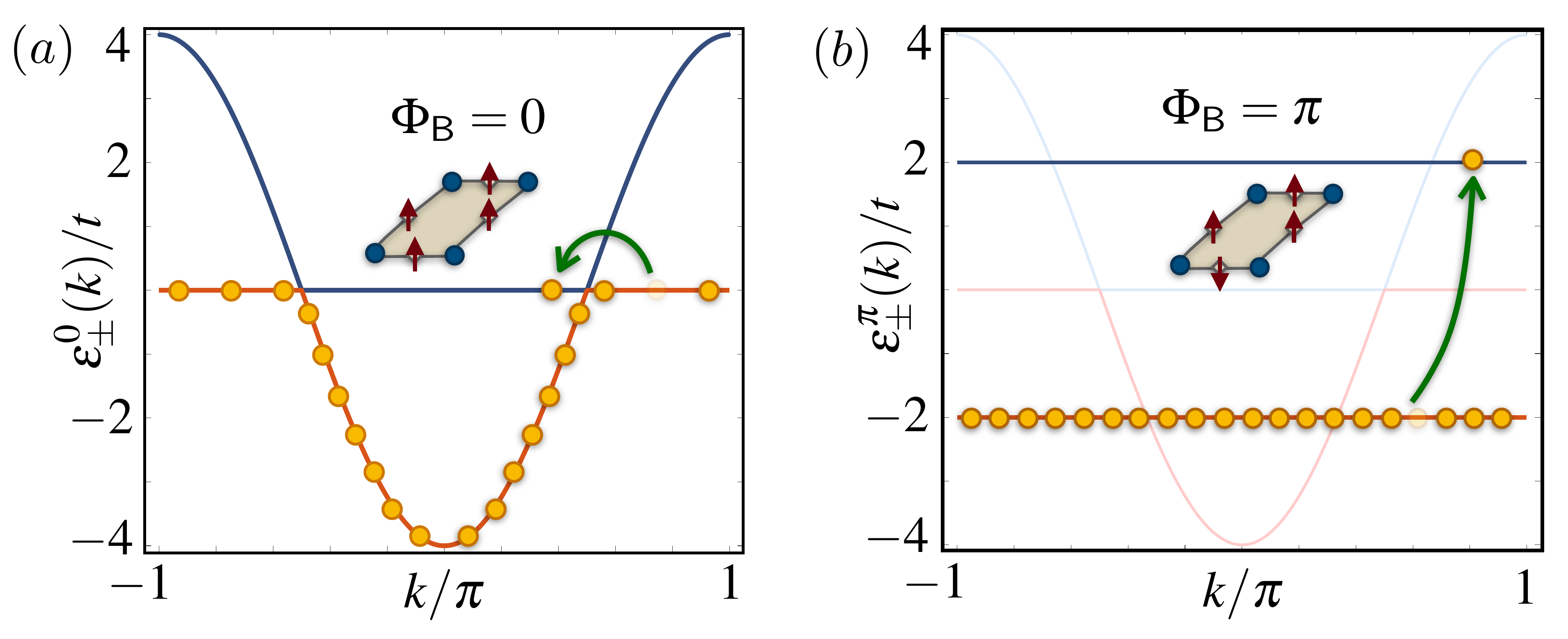}
\caption{\label{fig:instability} \textbf{Aharonov-Bohm instability:} Band structure of the Creutz-Ising gauge theory for $J=h=\Delta=0$, and corresponding filling of the fermionic sector. The green arrows depict the lowest-energy particle-hole excitations of the half-filled ladder. {\bf (a)} For an Ising background with a vanishing flux $\Phi_{\mathsf{B}}=0$, the ground-state corresponds to a gapless state with macroscopic degeneracy. {\bf (b)} For   $\Phi_{\mathsf{B}}=\pi$, there is destructive Aharonov-Bohm interference that opens a gap, forming two  flat bands, and  lowering the ground-state energy (the semi-transparent lines serve to compare  to the  $\Phi_{\mathsf{B}}=0$ case). }
\end{figure} 


\section{\bf Aharonov-Bohm Instability}
\label{sec:instability}

We start by exploring the limit of zero electric-field strength $h=0$. In the following, and unless stated otherwise, we fix $J = 0$. Here, the Ising fields have  vanishing   quantum fluctuations, and the  fermions tunnel in a classical $\mathbb{Z}_2$ background $\ket{\{\sigma_{(\boldsymbol{i},\boldsymbol{j})}\}}$, where  $\sigma_{(\boldsymbol{i},\boldsymbol{j})}=\pm1$ are  the eigenvalues of the $\sigma^z$ link operator. In this limit, there are only two translationally-invariant ground-states corresponding to the  $0$- or $\pi$-flux configurations, namely  $\langle\textsf{B}_{_{\boldsymbol{i}}\hspace{0.2ex}\plaquetteA}\hspace{1.5ex}\rangle=\langle \textsf{B}_{^{\boldsymbol{i}}\hspace{0.2ex}\plaquetteB}\hspace{1.5ex}\rangle=\pm1$. The fermions minimise their energy in these backgrounds by partially filling the corresponding energy bands $\epsilon_{\pm}^0(k)$ or $\epsilon_{\pm}^\pi(k)$.

For vanishing imbalance $\Delta=0$, and considering periodic boundary conditions,  these  band structures read
\beq
\epsilon_{\pm}^0(k)=-2t\cos k\pm2t|\cos k|,\hspace{3ex} \epsilon_{\pm}^\pi(k)=\pm 2t,
\eeq
where $k\in[-\pi,\pi)$. As depicted in Fig.~\ref{fig:instability}{\bf (a)}, for magnetic flux  $\Phi_{\mathsf{B}}=0$, the half-filled ground-state corresponds to a gapless state. Conversely, for  $\Phi_{\mathsf{B}}=\pi$  flux (Fig.~\ref{fig:instability}{\bf (b)}), the band structure consists of two flat bands, such that the half-filled ground-state is a single gapped state with a fully-occupied lowest band. By direct inspection of  Fig.~\ref{fig:instability}, it is apparent that the $\pi$-flux case is energetically favourable. This is indeed the case, as  one finds that $E_{\rm gs}^{\pi}=-2tN_{\rm s}<-(4t/\pi)N_{\rm s}=E_{\rm gs}^{0}$. Recalling  the Peierls instability in 1D metals~\cite{peierls}, where the underlying lattice adopts a dimerized configuration and  a gap is opened in the metallic band; here, it is  the Ising fields which adopt a $\pi$-flux configuration   leading to a gap opening in the fermionic sector. This spontaneous generation of a $\pi$-flux is in accordance with Lieb's result for bipartite lattices~\cite{PhysRevLett.73.2158} but, in contrast to the square lattice~\cite{Assaad_2016,Gazit_2017,Gazit_2018}, it does not lead to a semi-metallic phase with emergent    Dirac fermions~\cite{WEN1989641}. In this case, it is an insulator with  complete band flattening  caused by  destructive  Aharonov-Bohm  interference   at $\Phi_{\mathsf{B}}=\pi$~\cite{PhysRev.115.485}, which can result in many-body localization \cite{Kuno_2020}. Due to the remarkable similarities with the  Peierls effect, we  call this effect the Aharonov-Bohm instability. 

This flux instability is actually generic for any imbalance  $\Delta>0$, in spite of the fact that  the bands gain curvature.  In this case, the corresponding ground-state energies are 
\beq
\begin{split}
E_{\rm gs}^0&=-\frac{4t}{\pi}\bigg(\left(1+\xi^2\right)^{\!\!1/2}\mathsf{E}\big(\theta_0\big)\bigg)N_{\rm s},\\
E_{\rm gs}^\pi&=-\frac{2t}{\pi}\bigg(\left|1+\xi\right|\mathsf{E}\big(\theta_\pi\big)+\left|1-\xi\right|\mathsf{E}\big(\tilde{\theta}_\pi\big)\bigg)N_{\rm s},
\end{split}
\eeq
where we have introduced the parameters $\xi=\Delta/4t$, $\theta_0=1/(1+\xi^2)$, $\theta_\pi=4\xi/(1+\xi)^2$, and  $\tilde{\theta}_\pi=-4\xi/(1-\xi)^2$. Additionally, we have used  the complete elliptic integral of the second kind $\mathsf{E}(x)=\int_0^{\pi/2}\!{\rm d}\alpha\left(1-x\sin^2\alpha\right)\!\!^{1/2}$. Once again, one can readily confirm that $E_{\rm gs}^{\pi}<E_{\rm gs}^{0}$, such that it is energetically favourable for the ground-state to lie  in the $\pi$-flux phase, which is generally  gapped except for $\xi=1$, namely $\Delta=4t$.


\subsection{Emerging Wilson fermions and SPT phases}

By exploring the  imbalanced case at long wavelengths, we can understand the insulating $\pi$-flux phase from a different perspective. Rather than the massless Dirac fermions that emerge  in the square-lattice $\pi$-flux phase~\cite{Assaad_2016,Gazit_2017,Gazit_2018}, we  get the following long-wavelength dispersion around $k_{\pm}=\pm\pi/2$
\beq
\label{eq:dispersion_masses}
\epsilon^\pi_{\pm}(k_{\pm}+p)\approx\pm\sqrt{(m_{\pm}c^2)^2+(cp)^2},\hspace{3ex}m_{\pm}=(\xi\pm1)/2t
\eeq
where $c=2t$ is the propagation speed, and $m_{\pm}$ are two  mass parameters. Except for $\xi=1$, we  get two massive relativistic fermions characterised by a different mass, which are known as Wilson fermions in a LGT context~\cite{wilson_1977}.

 The fact that the Wilson masses are different $m_+\neq m_-$ turns out to be crucial in connection to the spontaneous generation of an SPT phase. The Chern-Simons form $\textsf{Q}^\pi_1=\frac{\ii}{2\pi}\bra{\epsilon_{-}^\pi(k)}\partial_k\ket{\epsilon_{-}^\pi(k)}{\rm d}k$~\cite{Ryu_2010}, equivalent to the Zak phase in 1D \cite{Zak_1989}, leads to a Chern-Simons invariant after integrating over all  occupied quasi-momenta
 \beq
 \label{eq:CS_1}
 \textsf{CS}^\pi_1=\int_{-\pi}^{\pi}\textsf{Q}^\pi_1=\frac{1}{4}\bigg({\rm sgn}(m_+)-{\rm sgn}(m_-)\bigg).
 \eeq
 One can define a gauge-invariant Wilson loop $ \mathsf{W}_1^\pi=\ee^{\ii2\pi \textsf{CS}^\pi_1}$ that detects the non-trivial topology when $ \mathsf{W}_1^\pi=-1$. This occurs when the pair of  Wilson fermions have masses with opposite signs. Accordingly, if $|\xi|<1$ (i.e. $-4t<\Delta<4t$), the topological Wilson loop is non trivial $ \mathsf{W}_1^\pi=-1$, and the emerging $\pi$-flux phase is an  insulating SPT phase.
 
 This result draws a further analogy between the Peierls and Aharonov-Bohm instabilities. In the former, when the instability  is triggered by an electron-lattice coupling that modulates the tunneling~\cite{PhysRevLett.42.1698, z2BHM,z2BHM_2}, one of the  dimerization patterns of the lattice leads to a non-zero topological invariant and an SPT phase~\cite{asboth_book}. In our case, there is no dimerization due to SSB since the local $\mathbb{Z}_2$ symmetry cannot be spontaneously broken. However, there are two gauge-invariant fluxes at $h=0$, and it is only  the $\pi$-flux configuration that leads to a non-zero topological invariant when $|\Delta|<4t$. We can thus conclude that, as a consequence of the Aharonov-Bohm instability, the fermions intertwine with the Ising fields in such a way that a gap is opened in the fermion sector with non-trivial topology.

\begin{figure}[t]
  \centering
  \includegraphics[width=1.0\linewidth]{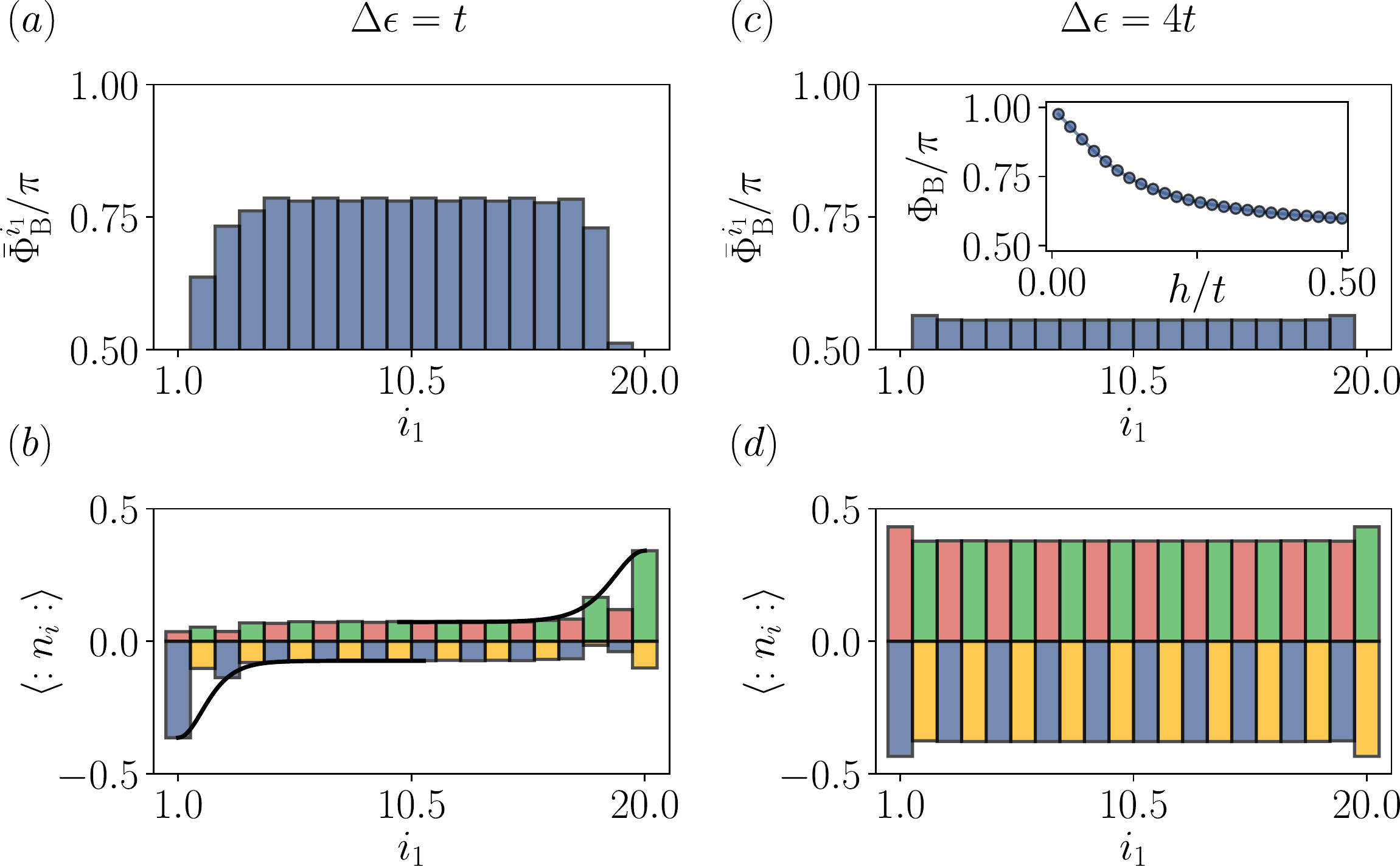} 
\caption{\label{fig:real_space} \textbf{Gauge-matter edge states: }  \textbf{(a)} Average Ising  flux $\bar{\Phi}_{\rm B}^{i_1}$ in the ground-state of the Creutz-Ising ladder  for $\Delta = t$ and $h = 0.1t$, for a ladder of length $N_s=20$ filled with $N=20$ particles. Due to the quantum fluctuations, the bulk flux  gets lowered $ \bar{\Phi}_{\rm B}^{\rm bulk} <\pi$. \textbf{(b)}  Fermionic occupation $\langle :n_{\boldsymbol{i}}:\rangle$ showing boundary peaks that can be identified with the topologically-protected edge modes. We use red and green to represent odd and even sites, respectively, of the lower leg, and blue and yellow for the upper one.  The solid black lines are obtained by fitting the latter to eq. \eqref{eq:edge_sech}, showing that the excess and defect of charge is due to the presence of edge states at zero energy.In \textbf{(c)} and \textbf{(d)} we can observe how, for a higher value of the imbalance ($\Delta = 4t$), the edge states disappear. The inset shows the flux in an infinite ladder as a function of $h/t$ for $\Delta = 0.1t$.}
\end{figure}


\subsection{Gauge-matter  edge states and  fractionalisation}

So far, our discussion has revolved around the zero electric field limit $h=0$, and assumed  periodic boundary conditions. From now onwards, we abandon this limit and explore the effect of quantum fluctuations in open Creutz-Ising  ladders (i.e. Dirichlet/hard-wall boundary conditions). Due to the bulk-boundary correspondence, when the bulk of the spontaneously-generated $\pi$-flux phase is characterised by a non-zero topological invariant~\eqref{eq:CS_1}, one expects that  edge states will appear at the boundaries of the ladder. In the context of LGTs, these states are lower-dimensional  domain-wall fermions~\cite{KAPLAN1992342} with the key difference that, in our case,  they are  generated via the Aharonov-Bohm instability. 

In Fig.~\ref{fig:real_space}, we show the real-space configuration of both  matter and Ising fields. We use a MPS-based algorithm~\cite{10.21468/SciPostPhysLectNotes.5} of the density-matrix renormalization group (DMRG)~\cite{PhysRevLett.69.2863}, setting the bond dimension to $D= 200$ for a ladder of leg length $N_{\rm s}= 20$ at half-filling, and introduce quantum fluctuations through  $h=0.1t$. In these figures, we display the Ising flux  
\beq
\label{eq:ising_flux}
\bar{\Phi}_{\textsf{B}}^{i_1}=\frac{\Phi_{\textsf{B}}^{_{\boldsymbol{i}}\hspace{0.2ex}\plaquetteA} + \Phi_{\textsf{B}}^{_{\boldsymbol{i}}\hspace{0.2ex}\plaquetteB}}{2}
\eeq
 averaged over the two trapezoidal plaquettes,  and the normal-ordered fermionic occupation 
 \beq
 \label{eq:fermionic_occupation}
 \langle:n_{\boldsymbol{i}}:\rangle=\langle c^\dagger_{\boldsymbol{i}}c^{\phantom{\dagger}}_{\boldsymbol{i}}\rangle-\rho,
 \eeq
  where $\rho=1/2$ at half-filling.  As shown in Fig.~\ref{fig:real_space}{\bf (a)} and {\bf (c)}, due to the quantum fluctuations, the Ising flux is no longer fixed at $\pi$. As the transverse field increases,  $\bar{\Phi}_{\textsf{B}}\to\pi/2$, which amounts to an electric-field dominated phase with a vanishing expectation values of the magnetic plaquettes $\langle\textsf{B}_{_{\boldsymbol{i}}\hspace{0.2ex}\plaquetteA}\hspace{1.5ex}\rangle=\langle \textsf{B}_{^{\boldsymbol{i}}\hspace{0.2ex}\plaquetteB}\hspace{1.5ex}\rangle=0$. In the inset we show how the flux $\Phi_{\mathsf{B}}$ changes with $h$ in an infinite ladder, where the ground state was obtained using the iDMRG algorithm with $D = 200$ \citep{10.21468/SciPostPhysLectNotes.5}. This change is continuous from $h=0$, suggesting that the flux-dominated phase found in the absence of quantum fluctuations with $\Phi_{\mathsf{B}}=\pi$ extends to finite values of $h$. In the next section, we will argue that this flux-dominated phase actually extends to the whole phase diagram, as in the case of $h=0$.
  
  In Fig.~\ref{fig:real_space}{\bf (b)}, we show that the corresponding fermion distribution is not translationally invariant, but displays an excess/deficit of charge around the boundaries of the ladder. This real-space distribution is  consistent with the existence of two topological edge states in the SPT phase, one of them being filled while the other one remains empty at half filling. We note that, in analogy with the phenomenon of charge-fractionalisation put forth by Jackiw and Rebbi~\cite{PhysRevD.13.3398}, when these zero modes are occupie/empty, an excess/deficit of $1/2$ fermion is formed around the boundaries. This fractionalisation can be readily observed in Fig.~\ref{fig:real_space}{\bf (b)}, where we also show  that the excess/defect of charge with respect to the bulk density on each leg of the ladder $\rho_{i_2}$ follows
\begin{equation}
\label{eq:edge_sech}
\langle n_{(j,i_2)}\rangle - \rho_{i_2} = \pm \frac{1}{4\xi_\ell}\text{sech}^2\left(\frac{j-j_\text{p}}{\xi_\ell}\right).
\end{equation}
Here, $j=2i_1$ (resp. $j=2i_1-1$) is the sublattice index for the lower (resp. upper) leg of the ladder, with $j_\text{p} = L$ (resp. $j_\text{p} = 0$), and $\xi_\ell$ is the  localization length of the corresponding edge state. This behaviour is a universal feature of zero modes in relativistic quantum field theories and  condensed-matter models~\cite{Campbell_1981}, and we show that it also holds for LGTs. In Fig.~\ref{fig:real_space}{\bf (b)} we observe how the edge states disappear for higher values of the imbalance $\Delta$, signaling a transition towards a trivial phase.

The presence of edge states points towards the robustness of the SPT flux-dominated phase described in the previous section, which thus persists as one introduces non-zero quantum fluctuations. Therefore, the SPT  phase should extend to a larger region in parameter space. Let us also highlight that, by looking at the enlarged  fluctuations of the Ising flux close to the boundaries (Fig.~\ref{fig:real_space}{\bf (a)}), one realises that  the edge states are indeed composite objects where both the matter and gauge degrees of freedom are intertwined. We will unveil a very interesting consequence of this intertwining below. 

\begin{figure}[t]
  \centering
  \includegraphics[width=1\linewidth]{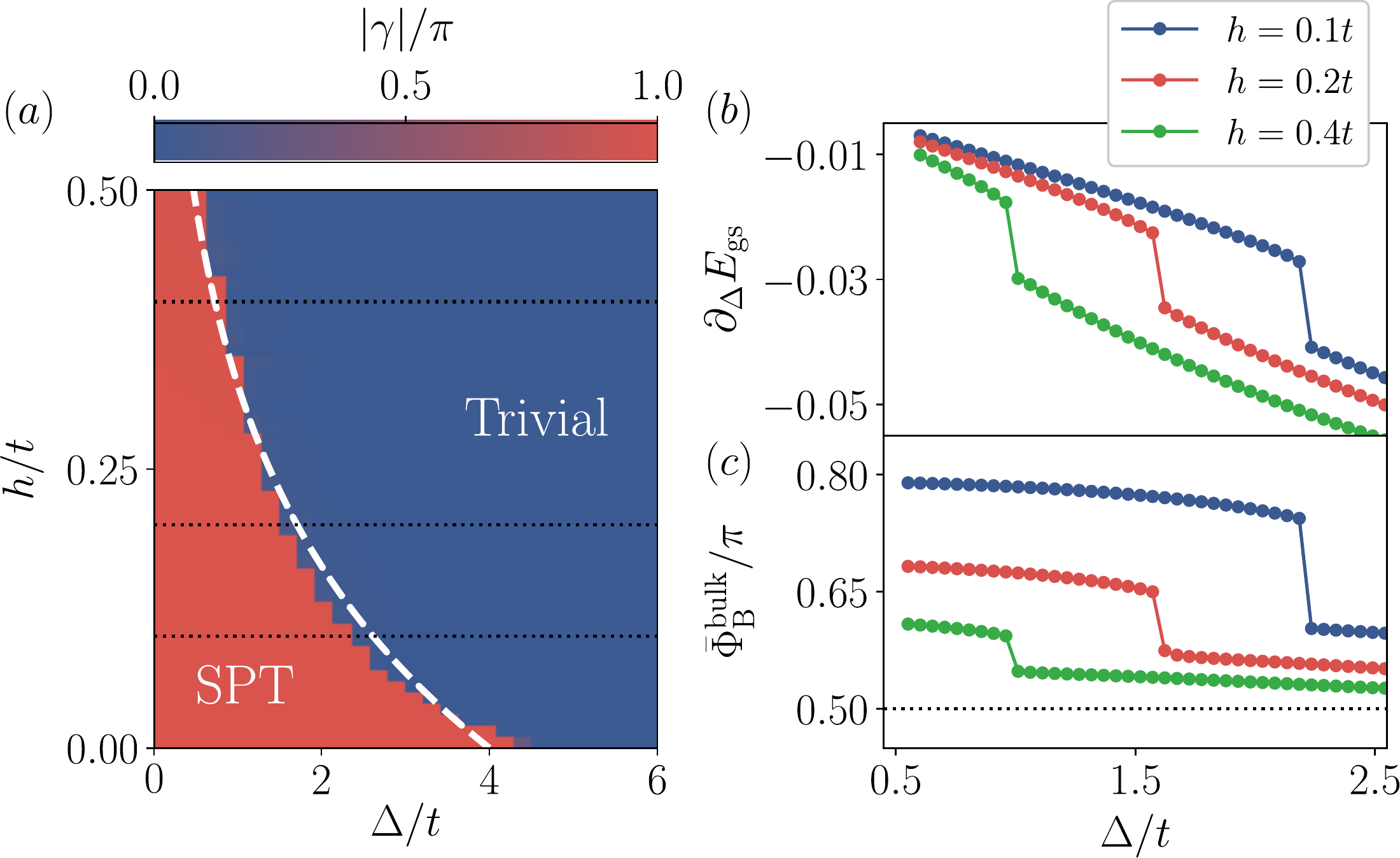} 
\caption{\label{fig:phase_diagram_h_eps} \textbf{Topological phase transitions: } \textbf{(a)} Contour plot of the many-body Berry phase, which allows differentiating between the SPT with $\gamma=\pi$ and the trivial band insulator (TBI) with $\gamma=0$. The dashed line is obtained by fitting the critical points to an exponential (see main text). \textbf{(b)} First derivative of the ground-state energy $\partial_{\Delta} E_{\rm gs}=(E_{\rm gs}(\Delta_1)-E_{\rm gs}(\Delta_2))/(\Delta_1-\Delta_2)$ with respect to the imbalance $\Delta/t$ for different values of $h/t$. \textbf{(c)} Average Ising flux $\Phi^{\mathrm{bulk}}_{\mathsf{B}}$ as a function of  the imbalance $\Delta/t$. The calculations were performed directly in the thermodynamic limit for a half-filled ladder.}
\end{figure} 


\subsection{Topological phase transitions}

We  explore  the extent of this SPT phase in parameter  space $(\Delta/t,h/t)$. The topological invariant~\eqref{eq:CS_1} is related to the  Berry phase $\gamma$ acquired by the ground-state $\ket{E_{\rm gs}(\theta)}$ along an adiabatic Hamiltonian cycle $\mathsf{H}(\theta)=\mathsf{H}(\theta+2\pi)$~\cite{berry_phase}, namely
\beq
\label{eq:berry}
\gamma=\ii\!\int_{0}^{2\pi}\!\!\!\!{\rm d}\theta\bra{E_{\rm gs}(\theta)}\partial_\theta\ket{E_{\rm gs}(\theta)}
\eeq
For non-interacting fermions in a classical $\mathbb{Z}_2$ background, one can use quasi-momentum as the adiabatic parameter $\theta=k$, such that $\gamma=2\pi\textsf{CS}^\pi_1$~\eqref{eq:CS_1}. However, 
as the electric field is switched on,  the Ising fields  fluctuate quantum-mechanically  mediating interactions between the fermions, and the quasi-momentum is no longer an appropriate adiabatic parameter. Building on ideas of   quantized charge pumping~\cite{pumping_thouless_int} and Hall conduction~\cite{PhysRevB.31.3372}, one can obtain a many-body Berry phase by  twisting the tunnelling $t\to t\ee^{\ii\theta}$ that connects the boundaries, and integrating over the twisting angle $\theta$. Interestingly, this concept can be generalized to systems with hard-wall boundary conditions~\cite{hatsugai_local}, since the twisting can actually be placed locally in any  link that respects the underlying symmetry that  protects the SPT phase, e.g. inversion symmetry in this case. 

We have computed the many-body Berry phase~\eqref{eq:berry}
for an infinite Creutz-Ising ladder using the iDMRG algorithm \citep{10.21468/SciPostPhysLectNotes.5}, yielding the phase diagram of Fig.~\ref{fig:phase_diagram_h_eps}{\bf (a)}. The 
 SPT phase is characterized by $\gamma=\pi$ in the red region, and is separated from a trivial band insulator (TBI) with $\gamma=0$ in the blue region by a   critical line that reaches  $\Delta\approx4t$ for $h=0$. This corroborates our previous interpretation~\eqref{eq:CS_1} in terms of the mass-inversion point of the emergent Wilson fermions at $\xi=\Delta/4t=1$. As the electric field $h$ increases, this inversion point flows towards smaller values of the imbalance   $\Delta$, which can be interpreted as a renormalization of the Wilson masses due to the interactions mediated by the gauge fields. 
 
 In this figure, we also show that   the numerical critical line can be fitted to an exponential $\xi_c = \xi_0{\rm exp}\{-h/h_\xi\}$, where  $h_\xi$ is a fitting parameter, and $\xi_0 = 1$ is fixed by  setting the critical point  at $\Delta/4t=1$ for $h = 0$. Let us remark that this exponential behaviour is consistent with the claim that the SPT phase and, in general, the Aharonov-Bohm instability and the flux-dominated phase, persists to arbitrarily-large values of the transverse $h$ when the imbalance is $\Delta = 0$. For zero imbalance, the appearance of the flat bands described previously endows the SPT phase with an intrinsic robustness to the  interactions mediated by the fluctuating gauge field.

Let us note that the critical line describes  first-order topological phase transitions, as can be appreciated in Fig.~\ref{fig:phase_diagram_h_eps}{\bf (b)}, where we display the  derivative of the ground-state energy $\partial_\Delta E_{~\rm gs}$ for three different values of electric field strength (dotted lines of Fig.~\ref{fig:phase_diagram_h_eps}{\bf (a)}). The discontinuous jumps   account for the first-order nature of the phase transitions. A similar discontinuity can be observed in the average magnetic flux~\eqref{eq:ising_flux}, evaluated at the bulk of the ladder (Fig.~\ref{fig:phase_diagram_h_eps}{\bf (c)}).


\section{\bf Topology from Connectivity}
\label{sec:topology}

The topological properties described in the previous section are associated, loosely speaking, to the matter degrees of freedom, since they are adiabatically connected to the static gauge field limit at $h = 0$\textemdash although matter and gauge are intertwined for any finite value of $h$. In this section, we focus on different topological effects that arise due to the dynamical nature of the gauge field. In particular, we provide quantitative evidence supporting the equivalence between the cross-linked ladder and a cylindrical geometry. This allows us to interpret our model as the thin-cylinder limit of a 2D IGT, and  to identify various topological properties such as the ground-state degeneracy or the presence of topological order throughout the whole phase diagram. We also show that the intertwining of the matter and gauge fields in the  SPT phase leads to a topological flux threading of the cylinder, and give further arguments for its survival to arbitrary  transverse fields. 

\subsection{The effective Creutz-Ising cylinder}

\begin{figure}[t]
  \centering
  \includegraphics[width=0.85\linewidth]{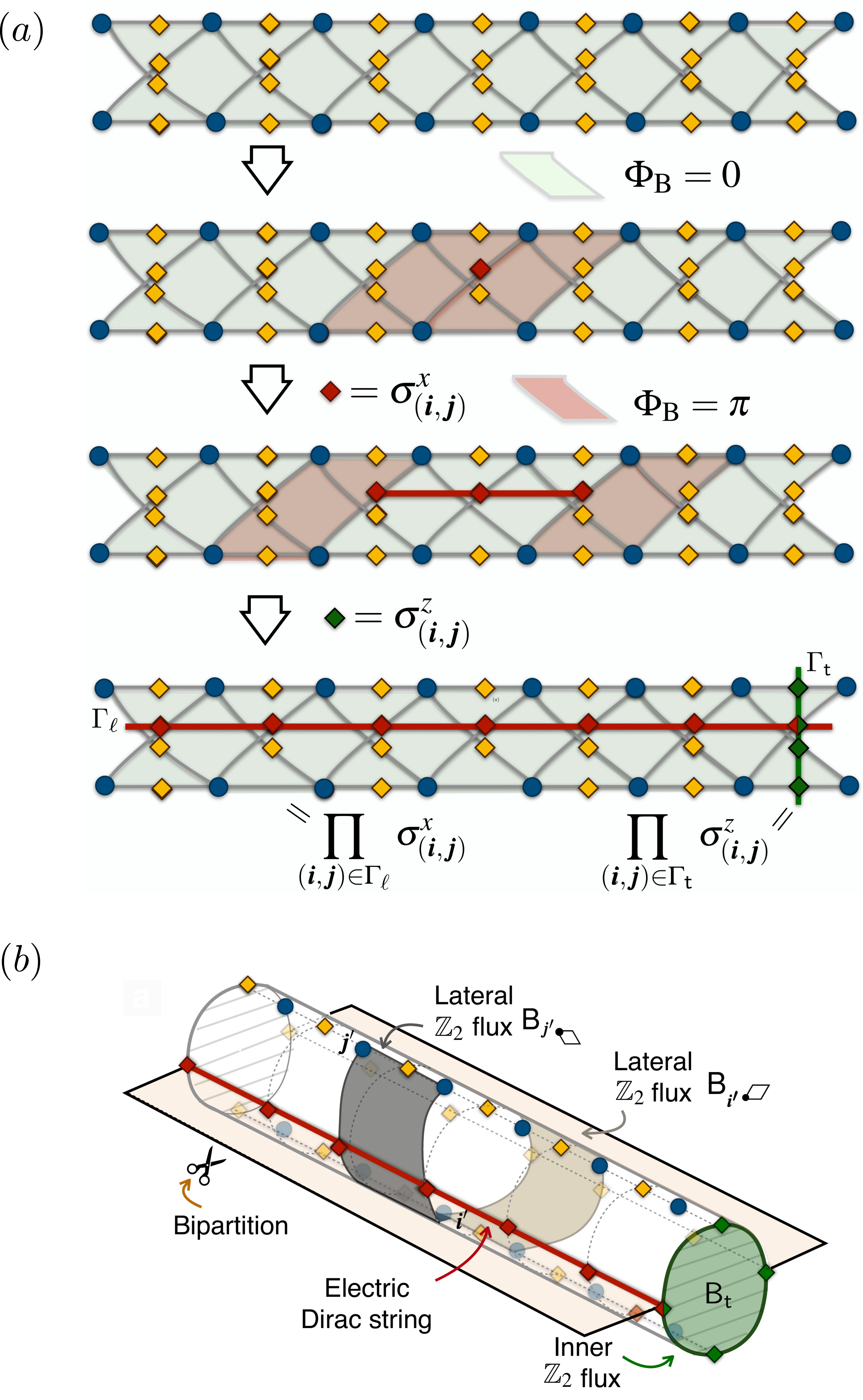}
  
\caption{\label{fig:scheme_dirac_string} \textbf{Ground-state degeneracy of the Creutz-Ising cylinder:} \textbf{(a)} In the $h/J\to0$ limit, the ground-state corresponds to the zero-flux state (green plaquettes), and excitations correspond to $\pi$-flux plaquettes (red) connected by a electric-field string (red line). By extending the string to the ladder edges, one recovers a different zero-flux ground-state state.    \textbf{(b)} By considering the crossed-link tunnelings of the ladder as two different paths enclosing an area, the Creutz ladder can be represented  as a thin-cylinder limit of a 2D LGT.}
\end{figure} 

Let us, momentarily, switch off the gauge-matter coupling and focus on the pure gauge theory  $\tilde{\mathsf{H}}_{\rm CI}(0,0,h,J)$ in Eq.~\eqref{eq:full_ham}. For $h/J\to0$ (with $J>0$), and for the sake of the argument, we assume that  $\ket{\rm g}$ is the single  ground-state  in a flux-dominated  phase with zero  flux per plaquette $\Phi_{\mathsf{B}}=0$. We note that the following argument, first applied to the IGT on a square ladder \cite{fradkin_2013}, is also valid for $\Phi_{\mathsf{B}}=\pi$ in the case $J<0$. As shown in Fig.~\ref{fig:scheme_dirac_string}\textbf{(a)}, flipping the Ising fields via $\sigma^x_{(\boldsymbol{i},\boldsymbol{j})}$ creates a pair of $\Phi_{\mathsf{B}}=\pi$ excitations at neighbouring plaquettes, which can be separated at the expense of flipping additional Ising fields along a  path $\Gamma_{\ell}$. By extending this path towards the boundaries of the ladder, the $\pi$ fluxes get expelled, and one recovers a state $\ket{\tilde{\rm g}}=\mathsf{D}_{\ell}\ket{\rm g}$ with vanishing flux  a  $\Phi_{\mathsf{B}}=0$, where 
\beq
\mathsf{D}_{\ell}=\prod_{(\boldsymbol{i},\boldsymbol{j})\in _{\Gamma_{\mathsf{L}}}}\sigma^x_{(\boldsymbol{i},\boldsymbol{j})}
\eeq
 is the so-called  Dirac string. Similarly to those in Eq.~\eqref{eq:four-body_flux}, one can define a  4-point correlator involving   Ising fields
\beq
\label{eq:inner_W}
\textsf{B}_{\mathsf{t}}=\!\!\prod_{(\boldsymbol{i},\boldsymbol{j})\in{\Gamma_{\mathsf{t}}}}\sigma^z_{(\boldsymbol{i},\boldsymbol{j})},
\eeq
where $\Gamma_{\mathsf{t}}$ is a vertical path that connects the two legs of the ladder. As  demonstrated below, $\Gamma_{\mathsf{t}}$ is equivalent to a path that wraps around a non-trivial cycle of a  cylinder, such that the correlator~\eqref{eq:inner_W} can be interpreted as a Wegner-Wilson loop operator measuring the flux threading the hole of the cylinder. 

In the lowest panel of Fig.~\ref{fig:scheme_dirac_string}\textbf{(a)}, one can see that  the Dirac string  shares only one common link with  the 4-point correlator, and thus anti-commutes   $\{\mathsf{D}_{\ell},\mathsf{B}_{\mathsf{t}}\}=0$. Conversely, the Dirac string   shares a pair of links with the trapezoidal plaquettes~\eqref{eq:four-body_flux}. and thus commutes with the  Hamiltonian $[\tilde{\mathsf{H}}_{\rm CI}(0,0,h,J),\mathsf{D}_{\ell}]=0$. As a consequence, if we assume that $\textsf{B}_{\mathsf{t}}|{\rm g}\rangle=+|{\rm g}\rangle$, we immediately obtain  $\mathsf{B}_{\mathsf{t}}|\tilde{\rm g}\rangle=-|\tilde{\rm g}\rangle$, whereas  $\tilde{\mathsf{H}}_{\rm CI}(0,0,h,J)|\tilde{\rm g}\rangle=E_{\rm gs}|\tilde{\rm g}\rangle$, $\tilde{\mathsf{H}}_{\rm CI}(0,0,h,J)|{\rm g}\rangle=E_{\rm gs}|{\rm g}\rangle$. Accordingly, the two  states are orthogonal and  have the same energy $E_{\rm gs}$. Our original assumption of a single ground-state thus needs to be dropped in favour of the existence of a two dimensional  ground-state manifold spanned by $\{\ket{\rm g}, |\tilde{\rm g}\rangle\}$.

The presence of such ground-state manifold can be a manifestation of topological order. 
 As outlined in the introduction, spin-liquid states with topological order can be characterised by a ground-state degeneracy that depends on the genus of the manifold in which they are defined~\cite{RevModPhys.89.041004}. The deconfined phase of the (2+1) IGT does indeed display this property, which is crucial in studies of  fractionalisation in high-$T_{\rm c}$ superconductors~\cite{senthil_2000,PhysRevB.63.134521}. In our case, the cross-linked tunnelings of Fig.~\ref{fig:scheme} can be understood as a planar projection of the two paths that traverse the different faces of a thin cylinder (see Fig.~\ref{fig:scheme_dirac_string}{\bf (b)}). From this perspective, the 4-point correlator~\eqref{eq:inner_W} is actually Wegner-Wilson loop that measured  the inner $\mathbb{Z}_2$ flux through the hole of the cylinder. As a consequence, the two-fold degeneracy follows from the non-trivial topology of the manifold: the two states that span the ground-state manifold $\ket{\rm g}, \ket{\tilde{\rm g}}$ have  a vanishing flux $\Phi_{\mathsf{B}}=0$ through the lateral surface of the cylinder, but only one of them $\ket{\tilde{\rm g}}$ has a $\pi$ flux through the cylinder's hole. This  ground-state $\ket{\tilde{\rm g}}$ is sometimes described as  a state with a vison (i.e. vortex excitation) trapped within the hole of the cylinder. 
 
 In a finite-size system, at any small but non-zero $h/J$, the ground-state degeneracy is lifted and the manifold splits into two eigenstates of $\mathsf{D}_{\ell}$ that amount to the symmetric and anti-symmetric superpositions of $\ket{\rm g}$ and  $\ket{\tilde{\rm g}}$. In the thermodynamic limit, the gap in the ground-state manifold closes exponentially, and any small perturbation tends to select one of the two eigenstates of $\textsf{B}_{\mathsf{t}}$, either  $\ket{\rm g}$ or $\ket{\tilde{\rm g}}$, which have a lower entanglement~\cite{zhang_2012}. In particular, we note that MPS simulations with finite bond dimension favour $\ket{{\rm g}}, \ket{{\tilde{\rm g}}}$ as ground-states with respect to any other choice. Having  less entanglement, their approximation for a fixed value of the bond dimension is more accurate, and thus their energy lower than the one of any other linear combination.  Below, we show that this effect  is also manifest in the presence of dynamical fermions, and that the specific intertwining of gauge and matter fields in the SPT phase induce well-defined magnetic fluxes inside the cylinder.

\begin{figure}[t]
  \centering
  \includegraphics[width=1\linewidth]{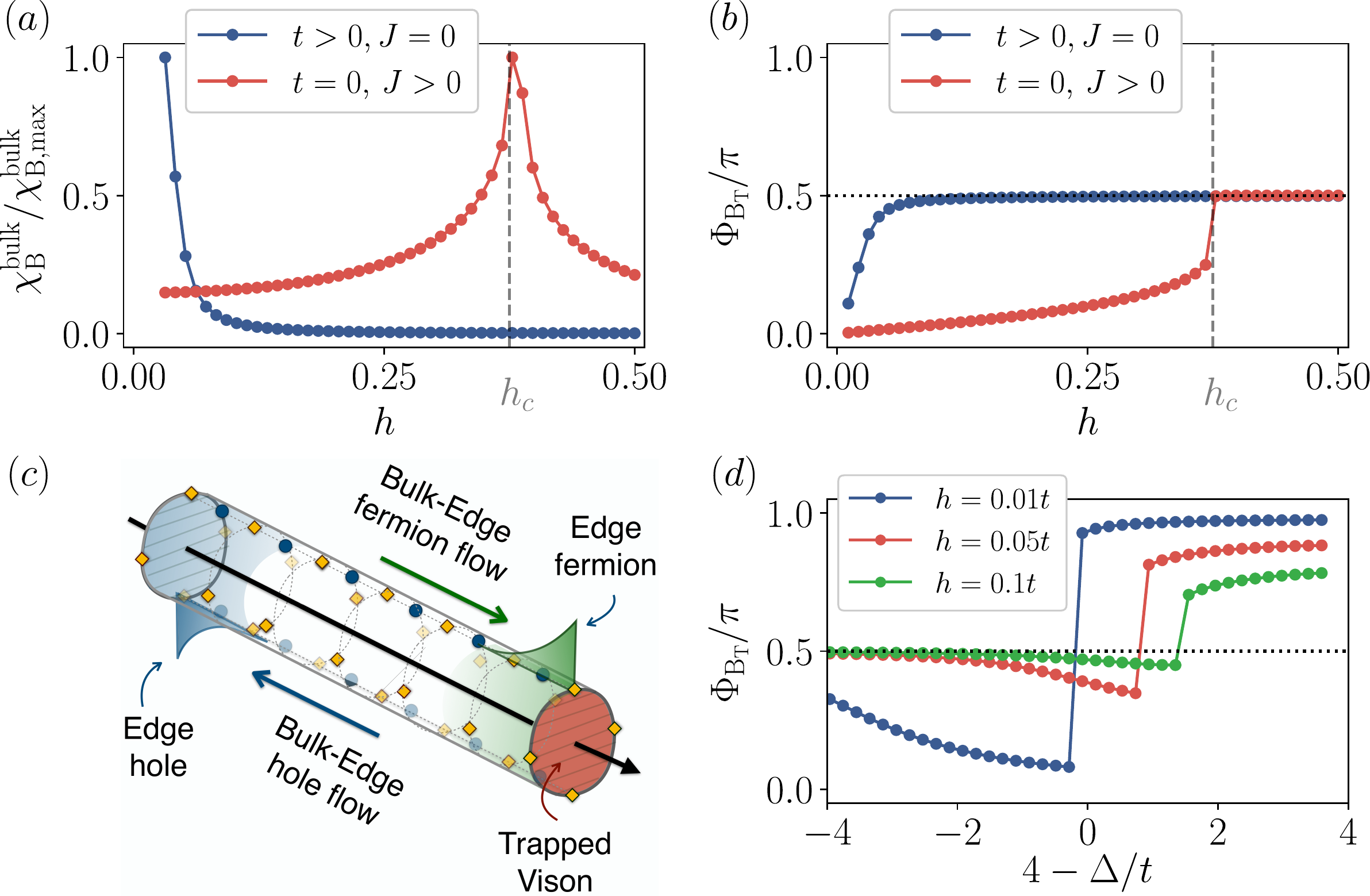}
\caption{\label{fig:fluxes} \textbf{$\mathbb{Z}_2$ magnetic fluxes in the infinite Creutz-Ising cylinder:} \textbf{(a)} $\mathbb{Z}_2$-flux susceptibility as a function of the electric field strength.  We compare the pure-gauge case, setting $t = 0$ and $J = 1$ (red circles) with the case in which the gauge fields interact with the  fermionic matter, setting $t=1$, $\Delta= 5t$ and $J=0$ (blue circles). In the first case, the susceptibility shows a diverging peak, signalling a quantum phase transition between deconfined and confined phases. In the presence of dynamical matter, however, there is no apparent divergence hinting at the absence of such a transition. \textbf{(b)} $\mathbb{Z}_2$ inner flux piercing the cylinder $\Phi_{\mathsf{B}_{\mathsf{t}}}$. In the pure-gauge case (red circles), this inner flux displays non-analytical behaviour across a  critical transverse field $h_{\rm c}$. For dynamical matter (blue circles), this tendency is not abrupt, and the inner  flux only attains the value $\Phi_{\mathsf{B}_{\mathsf{t}}}\approx\pi/2$ asymptotically without any non-analyticity. \textbf{(c)} Topological flux threading relating the existence of edge states to a trapped vison inside the cylinder. \textbf{(d)} Similarly to the Berry phase $\gamma$, the inner flux  $\Phi_{{\mathsf{B}}_{\mathsf{t}}}$   changes from $0$ to $\pi$ as one crosses the critical point separating TBI and SPT. The deviations from those precise  are due to quantum fluctuations.  }
\end{figure}

\subsection{Magnetic fluxes and Ising susceptibility}

We explore now the properties of the ground state as we departure from the limit $h/J\rightarrow 0$. For the pure-gauge case described above, in particular, a phase transition takes place at a finite value of $h/J$. As announced in Sec~\ref{sec:summary},  the  cross-linked ladder  geometry allows for  a confinement-deconfinement phase transition akin to the (2+1) IGT~\cite{Wegner_z2}. This phase transition can be probed by the $\mathbb{Z}_2$-flux susceptibility $\chi^{\rm bulk}_{\!\!\phantom{.}_{\mathsf{B}}}=\partial \bar{\Phi}_{\textsf{B}}^{\rm bulk}/\partial h$, evaluated through the magnetic flux~\eqref{eq:ising_flux} at the bulk of the ladder. It can be show that, after a duality transformation, the pure-gauge Ising gauge theory is equivalent to the quantum Ising model in a transverse field \cite{Wegner_z2}. In this picture, the $\mathbb{Z}_2$-flux susceptibility is equivalent to the susceptibility of the magnetization, which acts as an order parameter in the Ising model. We use iDMRG to obtain the approximation of the ground-state of the system defined on an infinitely-long ladder as an MPS. The system is thus equivalent to a $2\times \infty$ cylinder: the thin-cylinder limit of a 2+1 fermionic IGT. The maximum bond dimension we have used is $D=200$, testing that it is sufficient to achieve a good convergence. As clearly evidenced by the  iDMRG results of Fig.~\ref{fig:fluxes}{\bf (a)}(red circles) , there is a peak in the $\mathbb{Z}_2$ susceptibility, whose height  actually diverges with the ladder size at the critical coupling $\left. h/J\right|_{\rm c}$. In Fig.~\ref{fig:fluxes}{\bf (b)}, we plot the value of the inner flux $ \Phi_{{\textsf{B}}_{\mathsf{t}}}=\arccos(\langle\textsf{B}_{\mathsf{t}}\rangle)$ through the hole of the effective cylinder  as a function of the transverse field  $h$ (red circles). The plot shows that, in the $h/J\to0$ limit, the cylinder has zero inner flux $ \Phi_{{\mathsf{B}}_{\mathsf{t}}}=0$, and  the ground-state is  $\ket{g}$ as anticipated. By increasing  the transverse field, quantum fluctuations change the inner flux, which acts as a non-local order parameter for the transition to the confined phase  displaying  a non-analytical behaviour as we cross  $\left. h/J\right|_{\rm c}$.

Let us now switch on the gauge-matter coupling, and see how this picture gets modified by the inclusion of dynamical fermions governed by $\tilde{\mathsf{H}}_{\rm CI}(t,\Delta,h,0)$ in Eq.~\eqref{eq:full_ham}. First of all, we find that  there is no peak  in the   $\mathbb{Z}_2$ susceptibility for any value of $h$   (see blue circles of Fig.~\ref{fig:fluxes}{\bf (a)}), which suggests the absence of a phase transition. Furthermore, we plot the value of the inner flux  in Fig.~\ref{fig:fluxes}{\bf (b)}, which again attains the value $\Phi_{{\mathsf{B}}_{\mathsf{t}}}=0$ in the $h/t\to0$ limit (blue circles). The zero inner-flux state can be understood as the generalization of  the    $\ket{g}$ ground-state to a situation that encompasses dynamical fermions intertwining with the Ising fields. As neatly depicted, the $\mathbb{Z}_2$ flux   changes smoothly  from $\Phi_{\mathsf{B}_{\mathsf{t}}}=0\to\pi/2$ as the electric field strength is increased. Therefore, the absence of non-analyticities  again suggests that there is a single flux-dominated phase for arbitrary transverse fields. Although we did not show here that, for a cylindrical geometry and in the presence of fermionic matter, this phase exhibits a degenerate ground-state manifold, we will argue at the end of the section that this is indeed the case using the topological entanglement entropy. Moreover, in the last section we will demonstrate that, in this phase, fermionic matter is deconfined.

\subsection{Trapped Visons from  topological flux threading}

 As discussed above for the pure-gauge limit, topological order becomes manifest through the two-fold  ground-state degeneracy $\{\ket{{\rm g}}, \ket{{\tilde{\rm g}}}\}$, and the absence/presence of a trapped vison. Yet, in the previous section (see Fig.~\ref{fig:fluxes}{\bf (b)}), we have only found the dynamical-fermion generalisation of $\ket{g}$. As described in Sec.~\ref{sec:instability},  the Aharonov-Bohm instability can lead to an SPT ground-state  or to a trivial band insulator (see Fig.~\ref{fig:phase_diagram_h_eps}{\bf (a)}). We now discuss  the difference of the intertwining of the gauge and matter fields in these two  cases, and unveil a very interesting interplay between the topological degeneracy and the existence of edge states in the SPT phase.
 
 To understand this interplay, let us recall Laughlin's argument for the   quantum Hall effect~\cite{PhysRevB.23.5632},  which states that a single charge is transferred between the edges of a quantum Hall cylinder  when a magnetic flux quantum is threaded through its hole. In the Creutz-Ising ladder, one can move  from the TBI  onto the SPT ground-state by gradually decreasing the imbalance $\Delta$. As the system crosses the critical point,  topological edge states will appear at the boundaries of the ladder, which can be seen as the result of charge being transferred  from the bulk to the edges (see Fig.~\ref{fig:fluxes}{\bf (c)}).  In contrast to Laughlin's pumping, where it is the external variation of the flux  which leads to charge transport, here it is the transition into a topological phase and the associated charge transfer which should generate a non-vanishing $\mathbb{Z}_2$ inner flux. In Fig.~\ref{fig:fluxes}{\bf (d)}, we confirm this behavior, and show that the $\mathbb{Z}_2$ inner flux changes from  $\Phi_{{\mathsf{B}}_{\mathsf{t}}}\approx0$ (TBI) to $\Phi_{{\mathsf{B}}_{\mathsf{t}}}\approx\pi$ (SPT)   at fixed  $h=0.01t$. 
 
  This effect  can be  understood as a {\it topological flux threading}, where the existence of edge states gets intertwined with the trapping of a vison through the cylinder's hole, giving access to the dynamical-fermion generalization of $\ket{{\tilde{g}}}$. We note that this phenomenon cannot be observed with a background static  field,  such as  the magnetic field of the quantum Hall effect, but is instead characteristic  of  LGTs with fermionic matter,  unveiling an interesting interplay between the  Berry phase and the inner $\mathbb{Z}_2$  flux. This  offers  a  neat alternative to the numerical demonstration of the two-fold ground-state degeneracy, typically hindered by finite-size effects. As the quantum fluctuations are increased by raising $h$, we see that one tends smoothly to the electric-field dominated phase $\Phi_{{\mathsf{B}}_{\mathsf{t}}}=\pi/2$, but the first-order topological phase transition between  SPT and TBI, and the intertwining of the edge and vison states is still captured by the discontinuity of the inner flux.

\begin{figure}[t]
  \centering
  \includegraphics[width=1.0\linewidth]{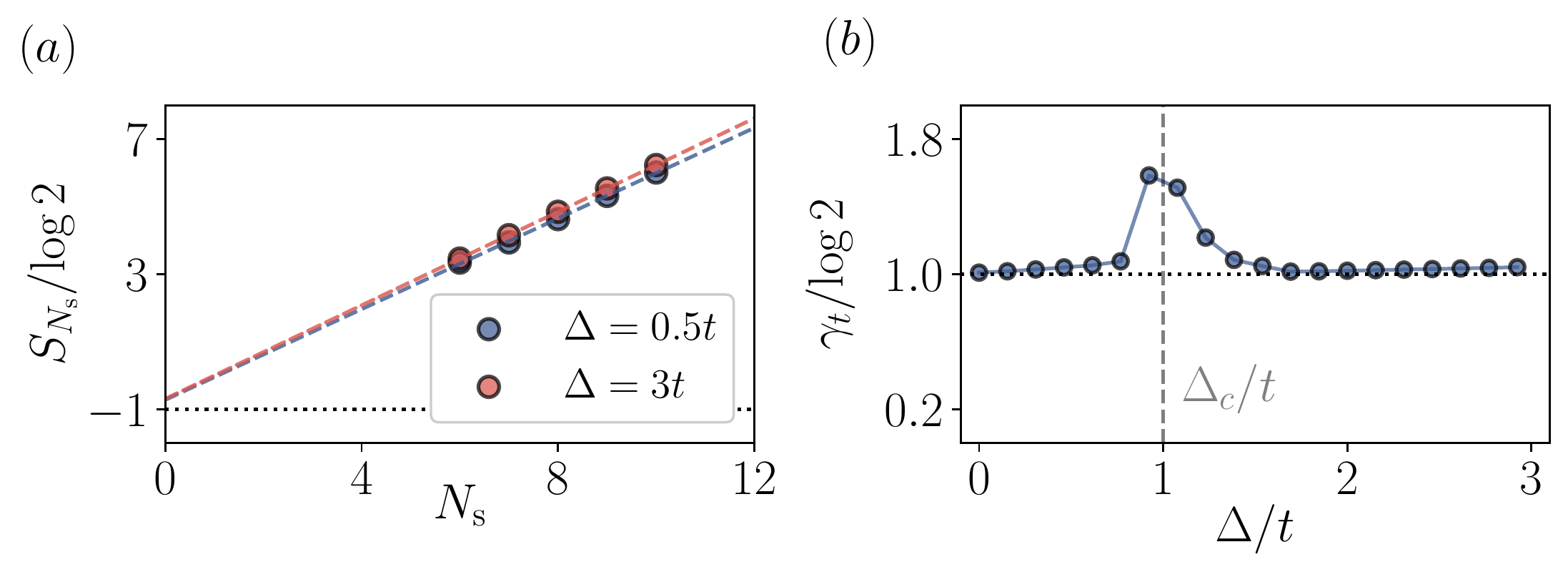}
\caption{\label{fig:top_entropy} \textbf{Topological entanglement  in the Creutz-Ising ladder: }  \textbf{(a)} Scaling of the entanglement entropy $S_{N_{\rm s}}$ with $N_{\rm s}$ setting  $h=0.02t$, and for different imbalances  corresponding to the SPT ($\Delta=0.5t$) and TBI ($\Delta=3t$) phases. In both cases, the topological correction to the entanglement entropy is $\gamma_{\rm t} = \mathrm{log}\,2$, signalling topological order. In \textbf{(b)} we represent this quantity as a function of $\Delta/t$. The grey dashed line denotes the position of the first-order transition for a finite cylinder of length $N_s = 10$.}
\end{figure}


\subsection{Topological entanglement entropy}

As argued in the previous section, the  ground-state degeneracy and the flux threading are  topological phenomena related to the underlying cylindrical  manifold. This raises the possibility that this quasi-1D IGT~\eqref{eq:full_ham} displays topological order, as occurs for Kitaev's toric code~\cite{kitaev_2003}. This is indeed the case in other quasi-1D geometries, such as the thin-torus limit of two-dimensional fractional quantum Hall states \cite{Bergholtz_2005, Grusdt_2014,Strinati_2017}. In recent years, quantum-information tools that quantify the entanglement of the ground-state have turned out to be extremely useful to characterise various many-body properties~\cite{LAFLORENCIE20161}. In particular, the   Von Neumann entanglement entropy  for a bi-partition of the ground-state $\ket{g}$ into two blocks $A$-$B$ of equal sizes is defined as $S(\rho_A)=-{\rm Tr}\{\rho_{A}\log\rho_{A}\}$, where  $\rho_{A}={\rm Tr}_{B}\{\ket{g}\!\!\bra{g}\}$ is the reduced density matrix. For a  (2+1)  topologically-ordered ground-state, this entanglement entropy scales as
\beq
\label{eq:ent_entropy}
S(\rho_{\rm r})=\alpha |\partial A|-\gamma_{\rm t},
\eeq
where $|\partial A|$ is the number of sites that belong to the boundary separating the $A$-$B$ regions,  $\alpha$ is a constant that characterises this entanglement area law, while $\gamma_{\rm t}$ is a universal sub-leading constant that quantifies the topological corrections~\cite{PhysRevLett.96.110404,PhysRevLett.96.110405}.  Although in a gapped phase the value of $\gamma_{\rm t}$ is constant, it has already been obseverd that close to a QPT there are strong finite size effects, and from numerical simulation it is very hard to extract a reliable determination of it \cite{tagliacozzo_2014}. Furthermore, the value of $\gamma_{\rm t}$ for bipartitions that are not contractible to a point depends on both the choice of the bipartion and the choice of the ground-state in the ground-state manifold \cite{zhang_2012}. 

In  order to reliably extract $\gamma_{\rm t}$, we turn to study finite-size Creutz-Ising ladders, interpreted through the mapping  to the thin cylinder of length $2\times N_{\rm s}$  of Fig.~\ref{fig:scheme_dirac_string}{\bf (b)}.  We consider a bipartition separating the two legs, such that $|\partial A|=N_{\rm s}$. In the effective manifold, this corresponds to a  longitudinal bipartition of the cylinder (see Fig.~\ref{fig:scheme_dirac_string}{\bf (b)}), such that the entropy~\eqref{eq:ent_entropy} should scale with the length of the cylinder. In this finite-size regime, the ground-state is an eigenstate of $\mathsf{D}_{\ell}$, and thus has minimal entropy.   $\gamma_{\rm t}$ should thus get saturated at its maximum value, namely  $\gamma_{\rm t}=\log(2)$. Our numerical analysis is limited to short ladders, as the particular bipartition limits the efficiency of the MPS routines. In Fig.~\ref{fig:top_entropy}{\bf (a)}, we plot the entanglement entropy as a function of the ladder length for two points deep in the SPT and the TBI. The fit of the data allows to confirm that $\gamma_{\rm t}\approx\log2$ in both the SPT and TBI phases. After repeating the same analysis for several values of the imbalance, we obtain Fig.~\ref{fig:top_entropy}{\bf (b)}. In this figure, $\gamma_{\rm t}$ is constantly very close to the expected $\log(2)$ within both SPT and TBI phases. It only departs significantly from that value  close to the phase transition, where the larger correlation length increases the finite-size effects \cite{tagliacozzo_2014}. The presence of a non-zero topological entropy is a further indication that  the complete gauge-matter system is topologically ordered both in the SPT phase and in the TBI. Due to  the lack of signature of  criticality  in our numerical results about the fluxes threading the cylinder and the bulk susceptibility of Figs.~\ref{fig:fluxes}{\bf (a)} and {\bf (b)} (blue circles), we  are thus confident  that the topologically-ordered phase survives for large values of the electric-field.

\begin{figure}[t]
  \centering
  \includegraphics[width=0.95\linewidth]{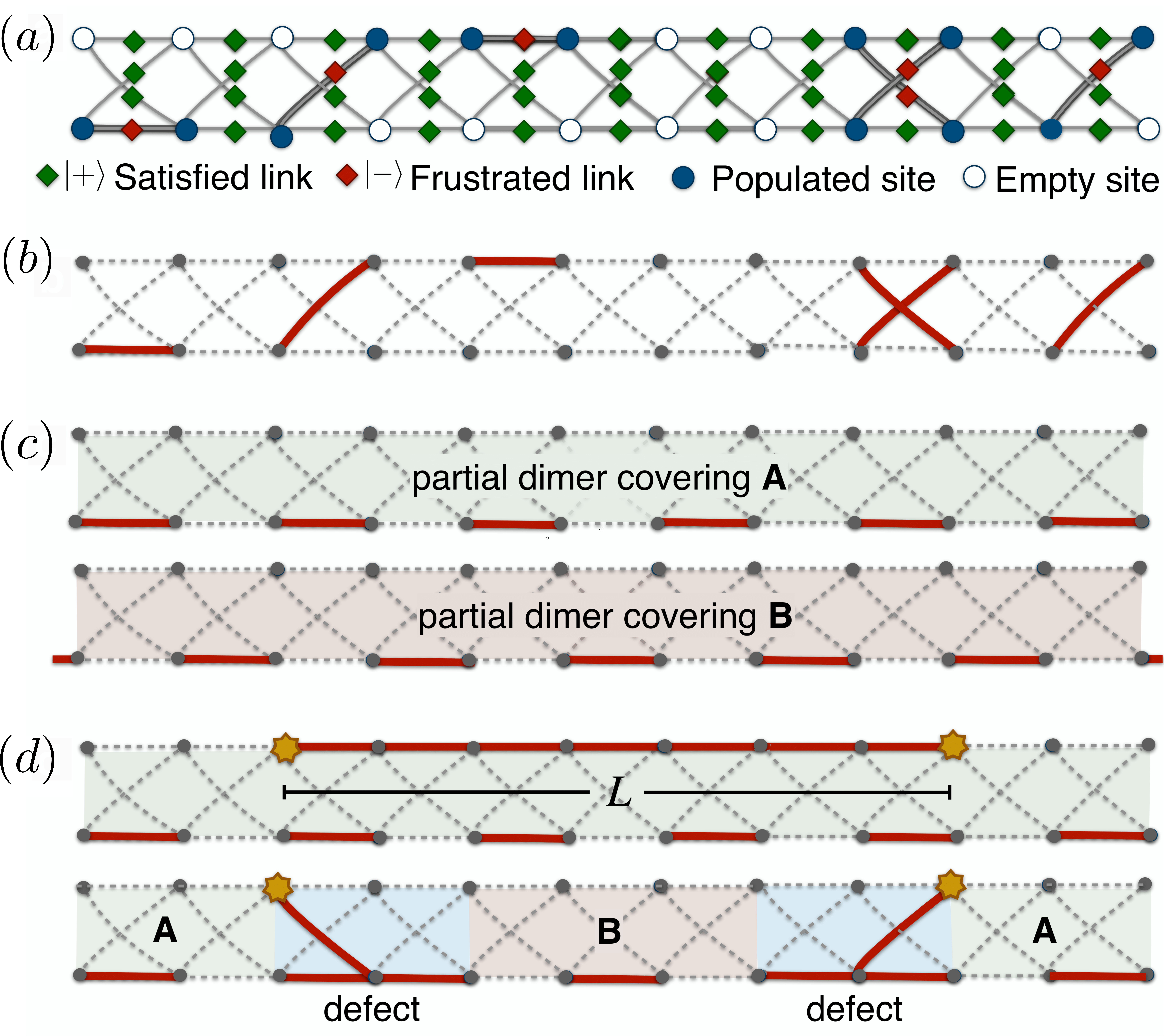}
\caption{\label{fig:scheme_deconfinement} \textbf{Dimer coverings and soliton deconfinement:} {\bf (a)} In the $h\gg t,J$ and $\Delta=0$  limit, Gauss law requires that some of the Ising fields will be frustrated due to the distribution of fermionic matter in the ladder. {\bf (b)}  The frustrated Ising fields can be identified with dimers (red bonds) that partially cover the lattice. {\bf (c)} Switching on $\Delta>0$ selects only the $A$ and $B$ dimer coverings. {\bf (d)} Adding  two  fermions changes the dimer covering, and it becomes energetically favourable to create topological defects in the dimer configurations, which can accommodate for the deconfined charges. }
\end{figure}

\begin{figure}[t]
  \centering
  \includegraphics[width=0.6\linewidth]{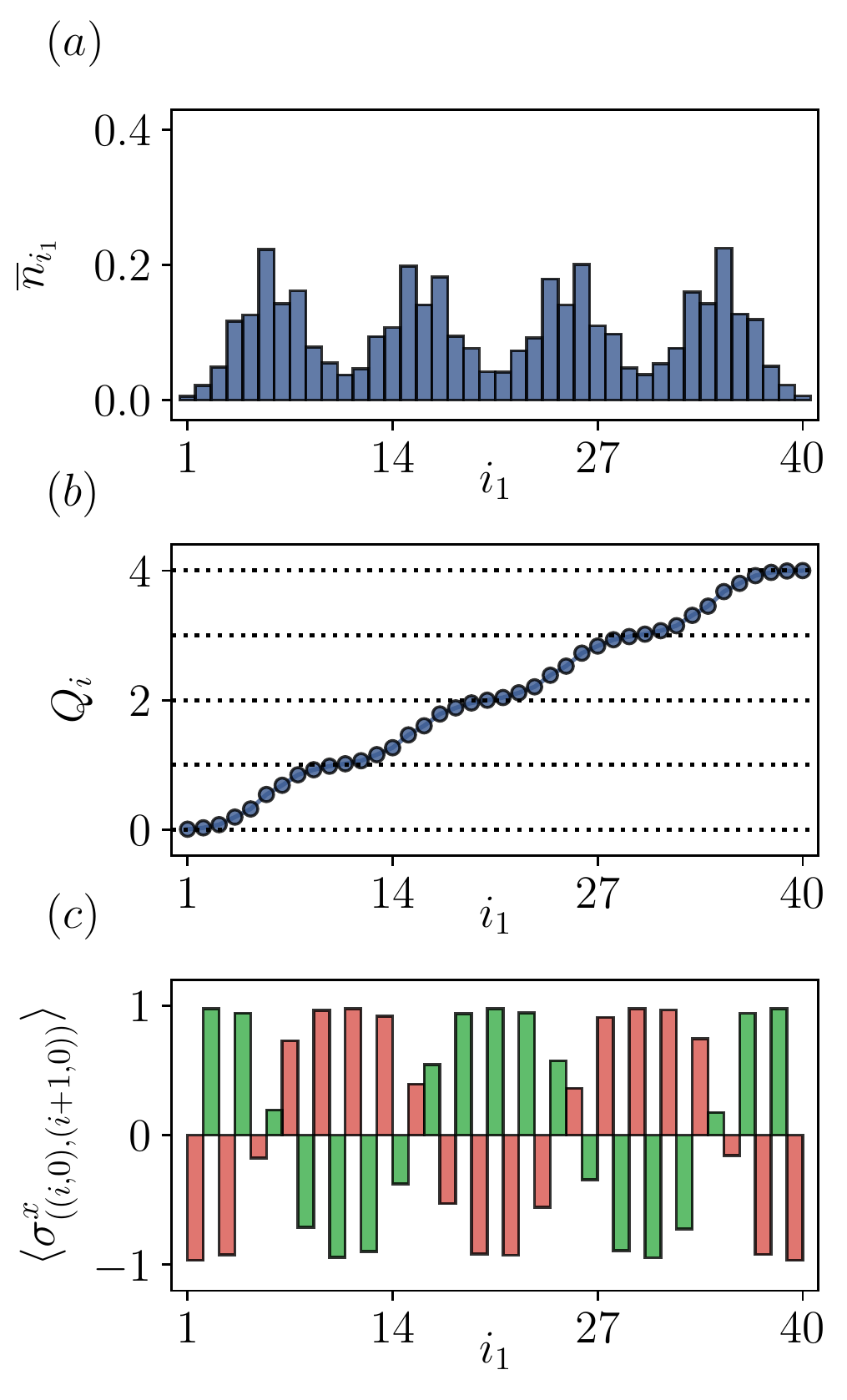}
\caption{\label{fig:deconf_dyn_charges} \textbf{Finite doping and soliton-induced deconfinement: } \textbf{(a)} Real-space fermionic occupation for a ladder of length $N_s = 40$ filled with $N = 44$ particles for $\Delta = 4t$,  $h = 0.2t$ and $J=0$, where the formation of a periodic crystalline structure can be appreciated. \textbf{(b)} Integrated charge $Q_i$ along the ladder, showing that each peak of excess charge with respect to half filling contains a fermionic number of one. \textbf{(c)} Corresponding electric field configuration $\langle \sigma^x_{((i,0),(i+1,0))}\rangle$ in the lower leg of the ladder, odd links are red and even are green. Topological solitons between the two degenerate electric field configurations appear, and the peaks of excess charge are located at the position of the defects.}
\end{figure}


\section{\bf $\mathbb{Z}_2$ Fermionic Deconfinement}
\label{sec:deconfinement}

In this section, we argue that the topologically-ordered flux-dominated phase described above shows fermionic deconfinement for any value of the transverse field. We first introduce the notion of gauge frustration, and how it generates deconfined topological defects when the system is doped above or below half filling. We then quantitatively characterise the absence of confinement using static charges, and we compare it with the more standard case involving string breaking. 

\subsection{Gauge frustration and topological defects}

Paralleling the situation in the standard (2+1) IGT~\cite{fradkin_2013}, the existence of topological order in the Creutz-Ising ladder suggests that the ground-state lies in a deconfined phase despite the lack of  plaquette interactions $J=0$. As outlined above, the absence of criticality for large $h$ suggests that this deconfinement may survive to arbitrarily-large electric-field strengths, which contrasts to the standard IGT~\cite{kogut_1979}. 

Let us start by discussing the   half-filled regime of the Hamiltonian~\eqref{eq:full_ham} for $h\gg t$ and $\Delta=0$. In this case, the link lsing fields  minimise their energy for $\ket{+}=(\ket{\uparrow}+\ket{\downarrow})/\sqrt{2}$. However, the presence of fermions can frustrate some links in order to satisfy   the constraints~\eqref{eq:gauss_law}, forcing the Ising fields to lie in $\ket{-}=(\ket{\uparrow}-\ket{\downarrow})/\sqrt{2}$. We call this   {\it gauge frustration}, namely the impossibility of simultaneously minimising all the individual Hamiltonian terms due to the Gauss constraint.   

 In contrast to pure gauge theories, this type of frustration can occur  in the even  sector  $q_i=0$, as some of the sites might be occupied by a dynamical fermion  (see Fig.~\ref{fig:scheme_deconfinement}{\bf (a)}). By  plotting only the frustrated links/bonds, one understands that the ground-state corresponds to  a  partial  covering of the ladder with  a single restriction: each site can  be touched by one bond at most (see Fig.~\ref{fig:scheme_deconfinement}{\bf (b)}). This is precisely the definition of a dimer, with the peculiarity that dimer models typically consider the complete covering of the lattice~\cite{PhysRev.124.1664,PhysRevLett.61.2376}, whereas in our half-filled case the ground-state will be a linear superposition of all partial dimer coverings. We note that, in the absence of dynamical fermions, the original connection of an IGT to a quantum dimer model in the large-$h$ limit was put forth by R. Moessner et {\it al.} by introducing a  static background $\mathbb{Z}_2$ charge $q_i=1$ at every site  (i.e. odd charge sector)~\cite{PhysRevB.65.024504}. In our case, the dynamical fermions allow for this dimer limit even in the absence of static charges, albeit  only  with partial coverings.

\begin{figure*}[t]
  \centering
  \includegraphics[width=1.0\linewidth]{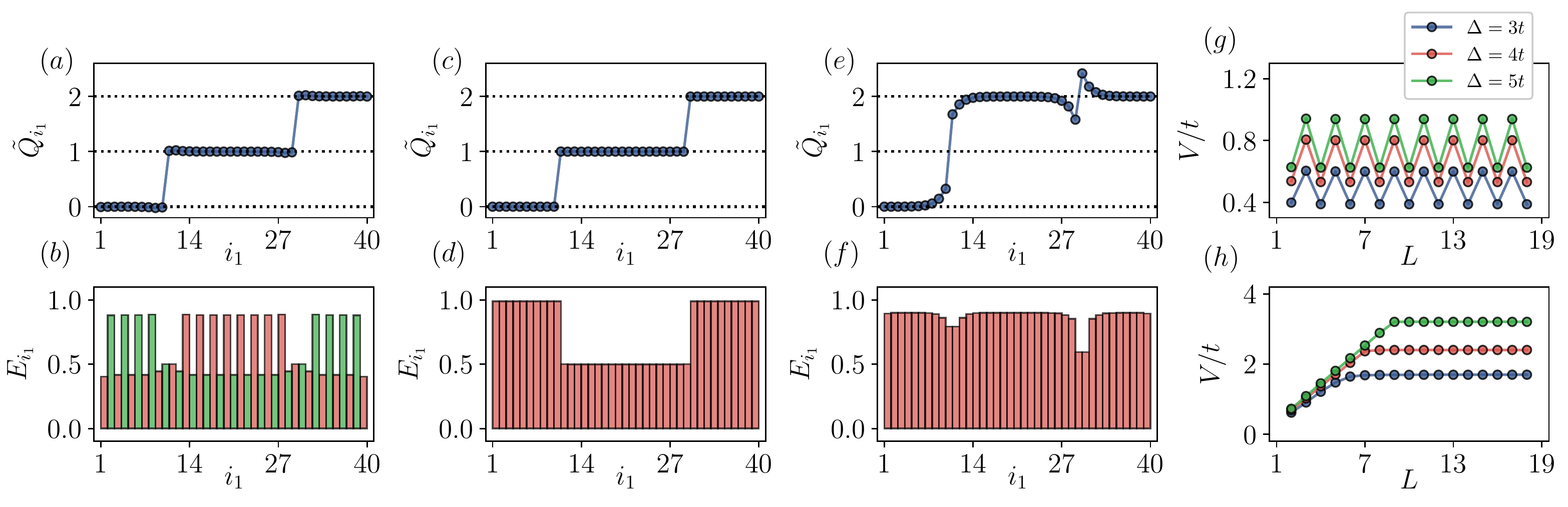}
\caption{\label{fig:deconf_static_charges} \textbf{Deconfinement versus  string breaking }: From \textbf{(a)} to \textbf{(f)} we represent the ground-state configuration corresponding to a ladder of length $N_{\rm s}=40$ at half filling with two extra static charges located in the upper leg and separated by $L=20$ sites, with $h = 0.2t$. \textbf{(a)} Integrated total charge $Q_i$ in the even charge sector for $\Delta = 3t$, showing two jumps of height one at the location of the static charges. \textbf{(b)} Each static charge creates a defect electric field to satisfy Gauss law. \textbf{(c)} Same as before but for  the imbalanced sector with $\Delta= 10t$. \textbf{(d)} In this case, an electric string develops between the charges. \textbf{(e)} and \textbf{(f)} depicts how, for $\Delta = 3t$ and in the imbalanced sector, a particle-antiparticle pair is created in the vacuum to screen the static charges, breaking the original string between them. \textbf{(g)} Potential energy $V/t$ between the two static charges as a function of the distance $x$ between them, for different values of $\Delta$, in the even  charge sector. \textbf{(h)} A similar calculation in the imbalanced sector.}
\end{figure*}

So far,  the imbalance has been fixed to zero. If we now allow for $\Delta>0$, the fermions will  preferably occupy  the lower leg, such that only two  degenerate coverings are relevant for the   large-$h$ ground-state (see Fig.~\ref{fig:scheme_deconfinement}{\bf (c)}). These two coverings, which we label as $A$ and $B$, are related  by a simple lattice translation and, yet, they  are essential for the deconfinement of the Creutz-Ising ladder.  If one adds a pair of fermions at a distance $L$  above half-filling, these must be accommodated in the upper leg, such that the Ising fields change to comply with the Gauss constraints. As depicted in Fig.~\ref{fig:scheme_deconfinement}{\bf (d)}, if we insist on maintaining one of the dimer coverings, say $A$, an electric field string must connect the fermions in the upper leg, such that the energy is $E(L)-E_0=hL$, and the charges are confined $V(L)\propto L$. This is the standard situation in the (2+1) IGT in the even sector~\cite{kogut_1979}. In our case, however, the two-fold coverings allow for a different situation: one can interpolate between the $A$ and $B$ configurations, such that $E(L)-E_0=3h$, and the charges are deconfined $V(L)\propto V_0$. The charges, which are no longer confined in pairs but localised at the topological soliton that interpolates between  $A$ and $B$, carry a non-zero $\mathbb{Z}_2$ charge, which is at the very heart of the notion of deconfinement. 

To assess the validity of these arguments, and  extend them beyond the $h\gg t$  limit, we explore the Creutz-Ising ladder  for finite doping  using the MPS numerics. In Fig.~\ref{fig:deconf_dyn_charges}{\bf (a)}, we depict  the  occupation  $\overline{n}_{i_1} = \langle:n_{(i_1,0)}:\rangle+ \langle:n_{(i_1,1)}:\rangle$ summed over the pair of sites in the upper and lower legs. This density displays an inversion-symmetric distribution of the extra doped charges, which are localised around distant centres  maximising their corresponding distances.  In Fig.~\ref{fig:deconf_dyn_charges}{\bf (b)}, we represent the integrated $\mathbb{Z}_2$ charge from the left boundary of the ladder $Q_{i_1}=\sum_{j_1\leq {i_1}}\overline{n}_{j_1}$. Comparing this profile to Fig.~\ref{fig:deconf_dyn_charges}{\bf (c)}, where we represent the electric-field configuration, it becomes manifest that 
each of the doped fermions is localised  within a topological soliton of the gauge fields. The bound fermion-soliton quasi-particles are deconfined, as they carry a unit $\mathbb{Z}_2$ charge (Fig.~\ref{fig:deconf_dyn_charges}{\bf (b)}), and can interact among each other forming a crystalline structure (Fig.~\ref{fig:deconf_dyn_charges}{\bf (a)}). To be best of our knowledge, our results confirm quantitatively this mechanism for the first time, and show that it can also appear in fermionic LGTs that combine topological order and SPT phases. We note that a similar deconfinement mechanism has  been suggested for the odd sector of a pure IGT in (1+1) dimensions~\cite{PhysRevB.65.024504}, based on   an analogy to the Peierls solitons in polymers~\cite{PhysRevLett.42.1698}. Our detailed analysis shows that, in our case,  this type of solitons characterised by  charge fractionalisation~\cite{PhysRevD.13.3398,z2BHM_4,z2BHM_3} are not the underlying mechanism explaining the deconfinement for the $h\gg t$ limit of IGTs. As dicussed above, the integrated charge around the solitons is quantitized in  units of the $\mathbb{Z}_2$ charge, but there is no signature of  charge fractionalisation (Fig.~\ref{fig:deconf_dyn_charges}{\bf (b)}). This result points to a different nature of topological defects in the magnetic- and electric-dominated phases, which will be the subject of detailed future studies.


\subsection{Deconfinement versus string breaking}

In this last section, we argue that the appearance of the  mechanism of soliton deconfinement depends on the particular charge sector~\eqref{eq:gauss_law}. So far, we have focused on the { even  sector}, which is characterised by the absence of  background $\mathbb{Z}_2$ charges $q_{\boldsymbol{i}}=0,\forall \boldsymbol{i}\in\mathbb{Z}_{N_{\rm s}}\times\mathbb{Z}_2$. We now make a full comparison with a different  sector, hereby referred to as  the {\it imbalanced sector}, where there is a static $\mathbb{Z}_2$ charge at each site of the lower leg, namely $q_{(i,0)}=1$,  $q_{(i,1)}=0$ $\forall i\in\mathbb{Z}_{N_{\rm s}}$. In this case, there are neither frustrated bonds in the ground-state, nor partial dimer coverings or solitons as in Fig.~\ref{fig:scheme_deconfinement}. Accordingly,  the situation and the confinement properties change completely. To quantify these differences, we  introduce two additional background charges on the upper leg of the ladder that are separated by a distance $L$, namely we add   $q_{(i_0,1)}=q_{(i_0+L,1)}=1$ to the two different charge sectors.

In Figs.~\ref{fig:deconf_static_charges}{\bf (a)}-{\bf (b)}, we represent the total integrated charge $\tilde{Q}_{i_1} = Q_{i_1}+\sum_{j_1<i_1}(q_{(j_1,0)}+q_{(j_1,1)})$,  which includes both the dynamical fermions and the background charges above the  even charge sector. We also depict  the underlying averaged electric field, $E_{i_1} = \sum_{i^{\vphantom{\dagger}}_2,i^\prime_2}\langle \sigma^x_{((i_1,i_2), (i_1+1, i^\prime_2))}\rangle/4$. The situation is analogous to the soliton-induced deconfinement discussed in the previous section, but the location of each solitons is now pinned to the position of the extra static charge, while in  Fig.~\ref{fig:deconf_dyn_charges} the unpinned solitons tend to maximise their spread and  distance forming a soliton lattice. In Figs.~\ref{fig:deconf_static_charges}{\bf (c)}-{\bf (d)}, we depict the same observables for the imbalanced sector. It is clear that an electric field line  connecting the static charges is established, which leads to the aforementioned confinement. The new aspect brought by the dynamical matter is that, by lowering the energy imbalance $\Delta$, it can become energetically favorable to create a particle-antiparticle pair that breaks this string and screens the static $\mathbb{Z}_2$ charges, as can be observed in Figs.~\ref{fig:deconf_static_charges} {\bf (e)}-{\bf (f)}. In these figures, one sees that the electric fields are restricted to the regions around the static charges, and that the $\mathbb{Z}_2$ charges are no longer unity, but get screened to $\tilde{Q}_i\,\mod2=0$, signaling the aforementioned string breaking.

We can make a quantitative study of the difference in the confinement properties of the two charge sectors by calculating the dependence of the effective potential with the distance between the static charges $V(L)$. In Fig.~\ref{fig:deconf_static_charges} {\bf (g)}, we present the results for the even charge sector and different values of $\Delta$.
In all cases, apart from an even-odd effect due to the so-called Peierls-Nabarro barriers associated with the defects~\cite{Peierls_1940,Nabarro_1947}, the energy does not grow with the distance, signaling deconfinement. In contrast, 
 in the imbalanced sector of Fig.~\ref{fig:deconf_static_charges} {\bf (h)},  the potential grows linearly with the distance  until the string breaks at a certain length, signalling confinement.
 

\section{\bf Conclusions and outlook}
\label{sec:conclusions}

In this work, we have identified novel topological effects in finite-density fermionic  gauge theories. In particular, we introduce a   minimal fermionic $\mathbb{Z}_2$ lattice gauge theory, the Creutz-Ising ladder, which allows  investigating the interplay between topology and gauge invariance. We show how, even in the absence of a plaquette term, the system presents a magnetic-flux dominated phase, in which a dynamical $\pi$ flux appears in the groundstate as a consequence of an Aharonov-Bohm instability. This phenomenon results from the interplay between gauge-invariant interactions and the particular connectivity of the model, which also gives rise to SPT phases in the fermionic sector. We characterize the properties of these phases, including the presence of protected gauge-matter edge states, through MPS-based numerical calculations, and use a topological invariant to find first-order phase transitions between the topological and trivial phases.

Our model can also be interpreted as a thin-cylinder limit of a (2+1) $\mathbb{Z}_2$ LGT. This equivalence allows us to uncover the presence of topological order by calculating the topological correction to the entanglement entropy. Topological order is also associated with the degeneracy of the ground-state, which can be characterised by two different fluxes threading the hole of the cylinder (i.e. presence or absence of a trapped vison). We have shown that, in the Creutz-Ising ladder, the topological order intertwines with topological symmetry protection, and this connection manifests in the change of the inner flux, and thus the trapping of a vison, when crossing phase transition lines towards the SPT phase such that edge states emerge from the bulk and localise within the ladder boundaries. This feature could facilitate the detection of topological order in future experiments.

Finally, we show how fermionic deconfinement, which accompanies the topologically-order phase, survives for the whole parameter space considered. This occurs due to the presence of deconfined topological defects associated to the fermionic quasi-particles, that appear on a frustrated background of electric fields imposed by gauge invariance. We investigate this mechanism using both static and dynamical charges, and compare it to the more standard confining case where string breaking usually takes place.

We believe that our results advance substantially the understanding of topological phenomena in lattice gauge theories. Moreover,  we have shown that the inclusion of dynamical fermions can stabilise a magnetic-dominated deconfined phase even in the absence of plaquette interactions. Therefore, our work identifies a new avenue for the realization of spin-liquid physics in LGTs, relevant for both condensed matter and high-energy physics, in cold-atom experiments based on state-of-the-art  building blocks that have already been used for quantum simulation purposes. This would allow to investigate complex phenomena, such as topological order and deconfinement, using minimal resources. The methods applied here can be used to further explore the static and dynamical properties of $\mathbb{Z}_2$ fermionic gauge theories, including the phase diagram at different fillings or the non-equilibrium  quench dynamics.

\section*{Acknowledgements}
We acknowledge interesting discussions with L. Barbiero, M. Di Liberto, N. Goldman, S. Hands, V. Kasper and E. Tirrito.
This project has received funding from the European Union's Horizon 2020 research and innovation programme under the Marie Sk\l{}odowska-Curie grant agreement No 665884, the Spanish Ministry MINECO (National Plan 15 Grant: FISICATEAMO No. FIS2016-79508-P, SEVERO OCHOA No. SEV-2015-0522, FPI), European Social Fund, Fundació Cellex, Generalitat de Catalunya (AGAUR Grant No. 2017 SGR 1341, CERCA/Program), ERC AdG NOQIA, EU FEDER, and the National Science Centre, Poland-Symfonia Grant No. 2016/20/W/ST4/00314. A.B. acknowledges support from the Ram\'on y Cajal program RYC-2016-20066,  CAM/FEDER Project S2018/TCS- 4342 (QUITEMADCM),  and PGC2018-099169-B-I00 (MCIU/AEI/FEDER, UE). L.T. acknowledges support from the Ram\'on y Cajal program RYC-2016-20594 and the ``Plan Nacional Generaci\'on de Conocimiento'' PGC2018-095862-B-C22.

\bibliography{bibliography_2}

\end{document}